\documentclass{SCIS2022}
\usepackage{amssymb,mathrsfs,cite,subfloat,booktabs,url,multirow,lmodern}

\begin{document}

\Year{2023}
\Month{}
\Vol{}
\No{}
\DOI{}
\ArtNo{}
\ReceiveDate{}
\ReviseDate{}
\AcceptDate{}
\OnlineDate{}

\title{Active Poisoning: Efficient Backdoor Attacks on Transfer Learning-Based Brain-Computer Interfaces}{Active Poisoning: Efficient Backdoor Attacks on Transfer Learning-Based Brain-Computer Interfaces}

\author[1]{Xue JIANG}{}
\author[1]{Lubin MENG}{}
\author[1]{Siyang LI}{}
\author[1,2]{Dongrui WU}{{drwu@hust.edu.cn}}

\AuthorMark{Jiang X}

\AuthorCitation{Jiang X, Meng L B, Li S Y, et al}

\address[1]{School of Artificial Intelligence and Automation, Huazhong University of Science and Technology, Wuhan {\rm 430074}, China}
\address[2]{Zhejiang Lab, Hangzhou {\rm 311121}, China}

\abstract{Transfer learning (TL) has been widely used in electroencephalogram (EEG)-based brain–computer interfaces (BCIs) for reducing calibration efforts. However, backdoor attacks could be introduced through TL. In such attacks, an attacker embeds a backdoor with a specific pattern into the machine learning model. As a result, the model will misclassify a test sample with the backdoor trigger into a prespecified class while still maintaining good performance on benign samples. Accordingly, this study explores backdoor attacks in the TL of EEG-based BCIs, where source-domain data are poisoned by a backdoor trigger and then used in TL. We propose several active poisoning approaches to select source-domain samples, which are most effective in embedding the backdoor pattern, to improve the attack success rate and efficiency. Experiments on four EEG datasets and three deep learning models demonstrate the effectiveness of the approaches.  To our knowledge, this is the first study about backdoor attacks on TL models in EEG-based BCIs. It exposes a serious security risk in BCIs, which should be immediately addressed.}

\keywords{Brain-computer interface, electroencephalogram, transfer learning, poisoning attack, backdoor attack}

\maketitle

\section{Introduction} \label{sect:Intro}

A brain-computer interface (BCI) uses human brain signals to directly interact with the computer~\cite{BCIIntro}. Electroencephalogram (EEG), which records electrical activities on the scalp of the brain, is the most widely used input signal in BCIs, due to its low cost and convenience~\cite{BCIReview}. Common paradigms of EEG-based BCIs include motor imagery (MI)~\cite{MI2001}, P300 evoked potentials~\cite{ZhouP300}, and steady-state visual evoked potentials (SSVEPs)~\cite{jin2021robust}.

An EEG-based BCI system usually consists of three parts: signal acquisition~\cite{Ji2022,Gu2021,chen2022}, signal analysis, and control action~\cite{Hao2021}. The signal analysis module is responsible for understanding the brain's intentions based on the collected brain signals. It generally includes signal processing~\cite{Makeig2012}, feature extraction~\cite{jin2019correlation,jin2021internal}, and pattern recognition~\cite{lotte2018review}. The latter two can be integrated into a single neural network if deep learning is used.

A major challenge in EEG signal analysis is that different subjects, or even the same subject in different sessions or tasks, have different neural responses to the same stimulus~\cite{saha2017evidence}. Therefore, it is difficult to build a generic model in EEG-based BCIs for different subjects, sessions or tasks~\cite{wu2020transfer}. In real-world applications, a calibration session is usually needed for a new subject to collect enough labeled data to tune model parameters~\cite{saha2017evidence}.

The calibration process in EEG-based BCIs is usually time-consuming and user-unfriendly, which greatly affects its real-world applications. Transfer learning (TL)~\cite{Pan2010} uses acquired data/knowledge in one or more source domains to improve the learning performance in a target domain, and hence it can be used to solve the above-mentioned problem. Specifically, in EEG-based BCIs, TL reduces the difference between a new subject/session/task (target domain) and existing subjects/sessions/tasks (source domains) to reduce or even completely eliminate the calibration needed for the target domain~\cite{wu2020transfer}. Many researchers have applied TL to EEG-based BCIs and achieved promising results~\cite{Jayaram2016,he2019transfer,he2020different}. In this study, we mainly consider cross-subject TL\footnote{Cross-subject transfer and cross-session transfer are essentially the same in TL. They are common in research and practice. Cross-task TL, where the label spaces of the source and target domains are different, is difficult and rarely studied in EEG-based BCIs. To our knowledge, only one work~\cite{he2020different} has considered cross-task TL in EEG-based BCIs so far.}.

Most existing BCI studies focused on improving their accuracy, but ignored their security. Recent studies have shown that EEG-based BCI systems are vulnerable to evasion attacks~\cite{zhang2019vulnerability,drwuUAP2021,zhang2020tiny} and backdoor attacks~\cite{Meng2020backdoor,bian2022SSVEP}. In a backdoor attack~\cite{gu2019badnets}, an attacker embeds a backdoor with a specific pattern into the machine learning model. For backdoor attacks in TL, the attacker can poison source-domain data to insert a backdoor. The resulting infected model will misclassify a test sample with a backdoor trigger into a prespecified class while still maintaining good performance on benign samples. Backdoor injection usually happens in the training phase when third-party data or models are used. In cross-subject TL, EEG data in the source domains are the main focus of the attacker, because he/she can offer such data for public downloading, but it is very difficult to notice or detect such backdoors. These backdoors would bring a critical security risk to the target subject, as pointed out in our previous research~\cite{zhang2019vulnerability}: ``\emph{EEG-based BCIs could be used to control wheelchairs or exoskeleton for the disabled, where adversarial attacks could make the wheelchair or exoskeleton malfunction. The consequence could range from merely user confusion and frustration, to significantly reducing the user's quality of life, and even to hurting the user by driving him/her into danger on purpose.}"

This study considers a new attack scenario in EEG-based BCI systems, in which a crafted trigger is added to the source data to create a dangerous backdoor in the TL model. A sample with the trigger (attacked sample) from a new subject will activate the backdoor and be classified into the target class\footnote{\emph{Target class} or \emph{target label} refers to the attacker-specified class in backdoor attacks, which is different from the concept of the target domain in TL.}. At the same time the benign samples will be normally classified.

\begin{figure*}[htpb] \centering
\includegraphics[width=.8\linewidth,clip]{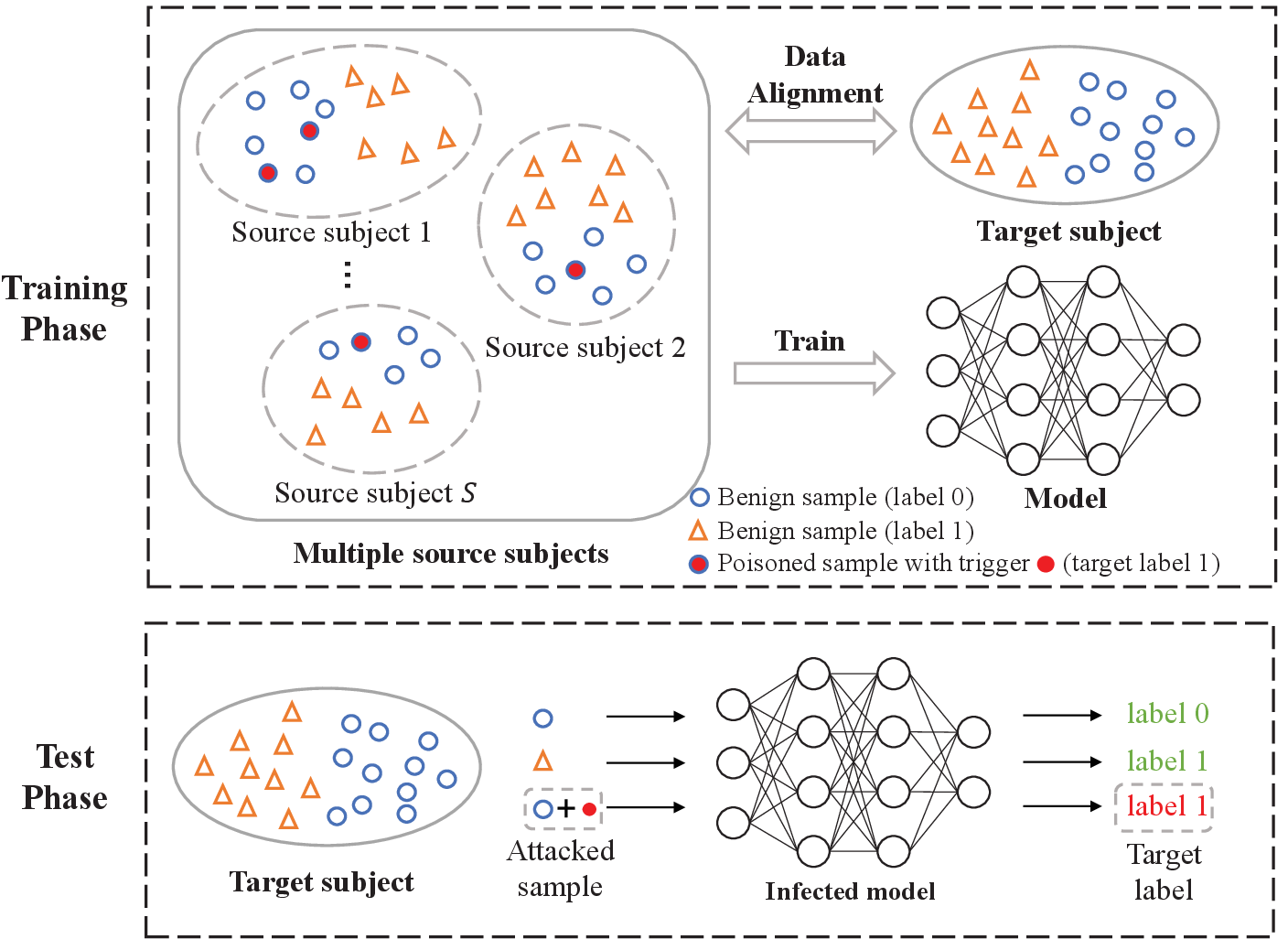}
\caption{Illustration of backdoor attacks in TL-based BCIs. Circles and triangles represent EEG samples from different classes. The red solid circle indicates the trigger specified by the attacker. In the training phase, the trigger is inserted into some source-domain samples to inject the backdoor, and their labels are modified to an attacker-specified class. Then, data alignment is used to make the data distributions from the source and target domains consistent. Finally, the target model is trained on the poisoned and aligned source-domain data. In the test phase, the classification of the benign samples is unaffected, but the attacked samples (with the backdoor trigger added) will be classified into the target class.} \label{fig:fig1}
\end{figure*}

Figure~\ref{fig:fig1} illustrates the idea of using data alignment-based unsupervised TL. Assume that the data of multiple source-domain subjects are obtained from a third party, some EEG samples in them have been inserted with a trigger (e.g., a narrow period pulse (NPP) or other types, as shown in Figure~\ref{fig:NPP} and Figure~\ref{fig:triggers}), and their labels are modified to the target class. An innocent user aligns the EEG data in the target domain with those in the poisoned source domains and then uses them together in TL for training a target domain model, completing the backdoor injection. In the test phase, the infected model will classify benign EEG samples in the target domain into its correct class, but it will misclassify those with the trigger into the target class specified by the attacker.

To improve the attack efficiency and stealthiness, we propose several active poisoning (AP) strategies to select samples in the source domains that are most beneficial to poison, including maximum diversity sampling (MDS), representativeness and diversity sampling (RDS), minimum uncertainty sampling (MUS), minimum model change sampling (MMCS), and their combinations. Experiments on four EEG datasets and three convolutional neural network (CNN) models validated that TL in EEG-based BCIs is vulnerable to backdoor attacks, and our proposed AP strategies can improve the attack success rate and stealthiness.

In summary, we make the following contributions:
\begin{enumerate}
  \item  Although TL is extensively used in EEG-based BCIs to reduce their calibration effort, its security has not been investigated in the literature. This is the first study to show that backdoor attacks could be performed on TL models in BCIs.
  \item We propose several AP strategies to optimally select source-domain samples to poison, making backdoor attacks efficient and stealthy.
  \item We show that backdoor attacks on TL models in EEG-based BCIs can achieve great success on four EEG datasets and three CNN models, even in very challenging scenarios, such as fine-tuning and data augmentation. We also verify that the AP strategies can greatly improve the attack success rate compared to traditional backdoor attacks under the same poisoning rate in different scenarios.
\end{enumerate}

The rest of the paper is structured as follows: Section~\ref{sect:RelatedWork} introduces related works on adversarial attacks in BCIs, TL, and active learning (AL). Section~\ref{sect:Approachology} proposes our backdoor attack scheme to TL in EEG-based BCIs and several AP strategies. Section~\ref{sect:Experiment} evaluates the attack performance of our proposed approach. Finally, Section~\ref{sect:Conclusion} draws conclusions.

\section{Related works}\label{sect:RelatedWork}

This section briefly reviews related works on adversarial attacks in BCIs, TL, and AL.

\subsection{Adversarial attacks in BCIs}

Recent studies have shown that machine learning models are vulnerable to adversarial attacks, posing great security risks. There are two main types of adversarial attacks: evasion attacks and poison attacks. Evasion attacks fool a machine learning model by adding imperceptible perturbations to a test sample. Poison attacks inject deliberately designed poisoned samples into the training set to manipulate the performance of a machine learning model. Backdoor attacks~\cite{gu2019badnets} are among the most dangerous poison attacks, where a secret backdoor is created in the model that allows the input sample with the backdoor trigger to be classified into a target class specified by the attacker.

Many adversarial attacks have been reported in image classification~\cite{Adversarialpatch}, speech recognition~\cite{AdvExamAudio}, and autonomous driving~\cite{qayyum2020securing}. In recent years, there have also been multiple studies on adversarial attacks in EEG-based BCIs. Zhang and Wu~\cite{zhang2019vulnerability} were the first to point out the existence of adversarial examples in EEG-based BCIs and performed white-box, gray-box, and black-box attacks on three popular CNN models in BCIs. Zhang \emph{et al.}~\cite{zhang2020tiny} showed that a tiny perturbation added to the EEG trial can mislead P300 and SSVEP spellers (which use traditional feature extraction and machine learning approaches) to output any character that the attacker wants. Meanwhile, Liu \emph{et al.}~\cite{drwuUAP2021} proposed universal adversarial perturbations for CNN classifiers in EEG-based BCIs. Both studies explicitly considered the causality in attacks. Recently, Bian \emph{et al.}~\cite{bian2022SSVEP} successfully attacked SSVEP-based BCIs using square-wave signals that are easy to generate and practically realizable.

All the above attacks on EEG-based BCIs are evasion attacks. Meng \emph{et al.}~\cite{Meng2020backdoor} were the first to study poison attacks in BCIs. They designed an NPP trigger for poisoning training data to embed a secret backdoor into the classifier. As a result, any test sample with the trigger will be misclassified into the target class. This study is different from \cite{Meng2020backdoor} in the following ways:
\begin{enumerate}
\item We consider a more practical TL scenario for backdoor attacks, where the poisoned source data could be provided for public downloading, and a benign user uses it to train a target model with an embedded backdoor. We also consider more challenging fine-tuning, data augmentation, and cross-task TL scenarios.
\item We propose several AP strategies for selecting poisoned samples, which can further improve the attack efficiency as compared with random selection in traditional backdoor attacks in various scenarios.
\item In addition to the NPP trigger proposed in~\cite{Meng2020backdoor}, we investigated other types of noise triggers and verified their effectiveness in AP attacks.
\end{enumerate}

\subsection{Adversarial attacks to TL}

Some researchers have studied adversarial attacks on TL models. Rezaei and Liu~\cite{rezaei2020target} implemented an evasion attack in re-training-based TL, in which the attacker does not need any information about the target model, except the publicly available pre-trained model, to generate adversarial examples which can be classified into any target category specified by the attacker. Wang \emph{et al.}~\cite{wang2018with} also proposed an evasion attack approach in re-training-based TL. They generated adversarial examples on a pre-trained model by minimizing the distance between the hidden-layer representations of the pre-trained model and the target model. Wang \emph{et al.}~\cite{wang2020backdoor} proposed a backdoor attack approach based on three commonly used backdoor attack defense approaches in re-training-based TL, by modifying the parameters of the pre-trained model to generate a robust backdoor that is difficult to defend against. Kurita \emph{et al.}~\cite{Kurita2020weight} also implemented backdoor attacks by modifying the parameters of the pre-trained model.

In summary, most attacks on TL so far were for re-training-based TL, and implemented by modifying the model parameters. Additionally, none of them considered EEG-based BCIs. To our knowledge, we are the first to perform backdoor attacks by modifying the source-domain data, rather than the model. Also, we are the first to study backdoor attacks to TL models in EEG-based BCIs.

\subsection{AL}

AL~\cite{Settles2009} is an effective approach to reduce the data labeling efforts, by optimally selecting the most useful instances to label. Many AL strategies, e.g., uncertainty sampling~\cite{sample2008analysis}, expected model change maximization~\cite{cai2017active}, and representativeness and diversity sampling~\cite{drwuSAL2019} have been proposed.

In the traditional data poisoning process of backdoor attacks, the trigger is placed in a number of clean samples which are randomly chosen from a training data set, enabling the model to learn a backdoor pattern. The next section introduces several AP strategies to expedite the backdoor learning. Different from the typical idea of AL that the selected samples are given to an oracle for labeling, in AP the selected samples are labeled into the target class, regardless of their true class.

\section{AP for backdoor attacks to TL in BCIs} \label{sect:Approachology}

This section introduces the attack scenario in TL-based BCIs, the backdoor trigger, and our proposed AP strategies for poisoned sample selection.

\subsection{Attack scenario}

This work aims to attack the TL model in BCIs. Assume that the attacker can poison the source-domain data, which is possible in practice. For example, the attacker can collect some benign EEG data, add poisoned samples, and then offer them for downloading. In downloading such data, the user usually needs to fill a form to indicate his/her affiliation, so the attackers know who is using the poisoned data.

We assume a data alignment based offline TL is used in EEG-based BCIs. Assume also there are $S$ source subjects, and the $s$-th source subject has $N_s$ labeled EEG samples $\{(X_s^n,y_s^n)\}_{n=1}^{N_s}$, where $X_s^n\in\mathbb{R}^{C\times T}$ is the $n$-th EEG sample and $y_s^n$ the corresponding label, in which $C$ is the number of EEG channels and $T$ the number of time domain samples. For binary classification, the label for $X_s^n$ is $y_s^n\in\{0, 1\}$. The target subject has $N_t$ unlabeled samples $\{X_t^n\}_{n=1}^{N_t}$.

The attacker selects $P$ non-target samples (i.e., Class $0$) from the data of the source subjects randomly or using an AP strategy described below, adds the crafted trigger $\mathbf{x}^*$ to each channel of them, and changes their labels to the target class (i.e., Class $1$) to obtain the poisoned samples $(\widetilde{X}_s^p,1)_{p=0}^{P_s}$, where $P_1+P_2+...+P_S=P$. Once the poisoned-source data are used in TL, the backdoor is automatically inserted.

For a new subject, the attacker injects the same backdoor trigger as in the source domain into the EEG sample of the target domain. Then, the samples with the backdoor trigger will be classified into the target class specified by the attacker, and all benign samples will be classified normally. In practice, using a simple, periodic and phase-independent trigger is more realizable for attacking an online BCI system.

\subsection{Trigger design}

We apply the NPP trigger proposed by Meng \emph{et al.}~\cite{Meng2020backdoor} in our backdoor attacks, because it is easy to implement and practically feasible, e.g., NPP as a type of common interference noise can be injected into EEG signals during data acquisition.

A continuous NPP can be determined by an amplitude $a$, a period $T$, a phase $\phi$, and a duty cycle $d$, i.e.,
\begin{eqnarray}
	\mathcal{N}_c(t) = \left\{
             \begin{array}{ll}
             0, &  nT\leq t < nT+\phi\\
             a, &  nT+\phi\leq t < nT+dT+\phi\\
             0, &  nT+dT+\phi\leq t < (n+1)T
             \end{array}
             \right..	
\end{eqnarray}

We set a random $\phi$ for each attacked samples to reduce the dependency of the backdoor on the phase, so that the backdoor attack performance is insensitive to injection time.

After discretization with a sampling rate $f_s$, the NPP can be expressed as
\begin{eqnarray}
	\mathcal{N}_d(i) = \left\{
             \begin{array}{lr}
             0, &  nTf_s\leq i < (nT+\phi)f_s\\
             a, &  (nT+\phi)f_s\leq i < (nT+dT+\phi)f_s\\
             0, &  (nT+dT+\phi)f_s\leq i < (n+1)Tf_s
             \end{array}
             \right..	
\end{eqnarray}

The discrete NPP was used as the trigger $\mathbf{x}^*\in\mathbb{R}^{1\times T}$. Given an EEG sample $X^n\in\mathbb{R}^{C\times T}$, the poisoned sample $\widetilde{X}^n$ can be obtained by adding the trigger to all channels simultaneously:
\begin{eqnarray}
\widetilde{X}^n=X^n + \mathbf{1}\cdot\mathbf{x}^*,
\end{eqnarray}
where $\mathbf{1}\in\mathbb{R}^{C\times1}$ is a column vector with all ones.

Compared with artifacts in EEG signals, the backdoor triggers have higher requirements on controllability and repeatability. In addition to NPPs, other possible triggers are discussed in Section~\ref{sect:trigger}.

\subsection{Active poisoning} \label{sec:AP}

Intuitively, if there are more poisoned samples in the source domain, then it is easier to learn the trigger pattern and perform backdoor attacks. However, poisoning too many samples reduces the stealthiness of backdoor attacks. It is desirable to select a small number of samples that are most effective in embedding the trigger.

AL aims to select a few most useful unlabeled samples to label for training, so that good learning performance can be achieved from a small number of labeled samples. Inspired by AL, we propose several AP strategies below.

\subsubsection{Maximum diversity sampling (MDS)}

Diversity-based AL approaches, e.g., GSx, GSy and iGS~\cite{wu2019active}, query the most scattered samples to label, so that a good global model can be learned. In backdoor attacks, diversity is also an important criterion for selecting the poisoned samples: if the poisoned samples are distributed in the entire input and/or output space, then the trigger is more universally embedded into the model, and hence it is easier to perform backdoor attacks.

\begin{figure}[htbp]\centering
\subfloat[]{\label{fig:fig2_a} \includegraphics[width=.4\linewidth,clip]{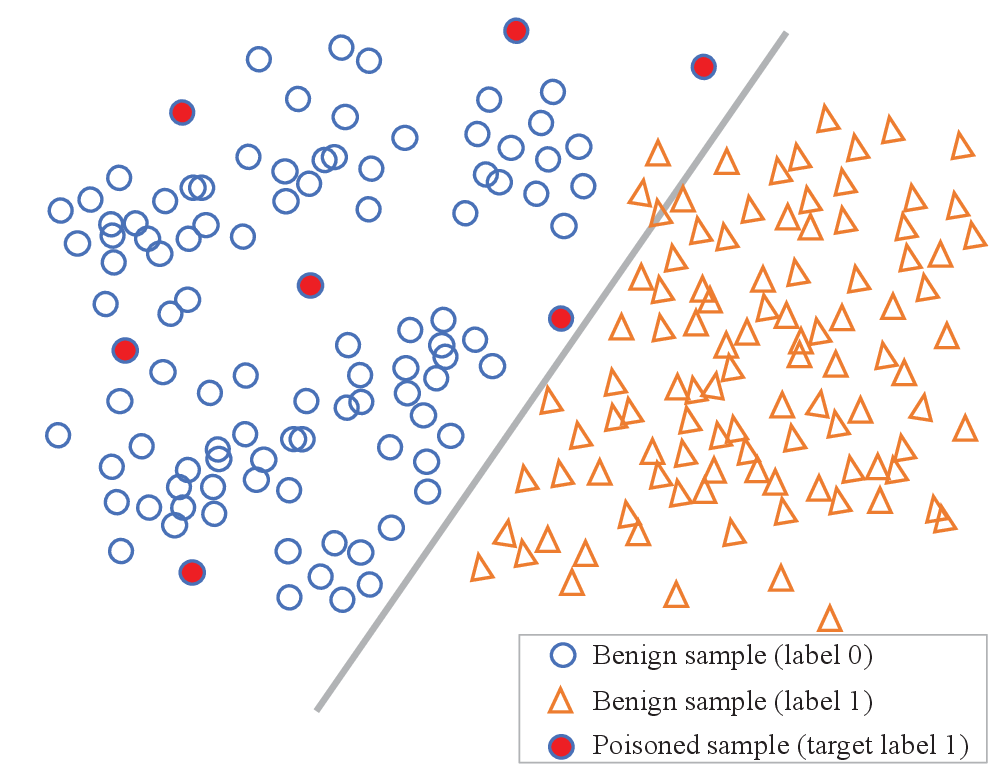}}
\quad
\subfloat[]{\label{fig:fig2_b} \includegraphics[width=.4\linewidth,clip]{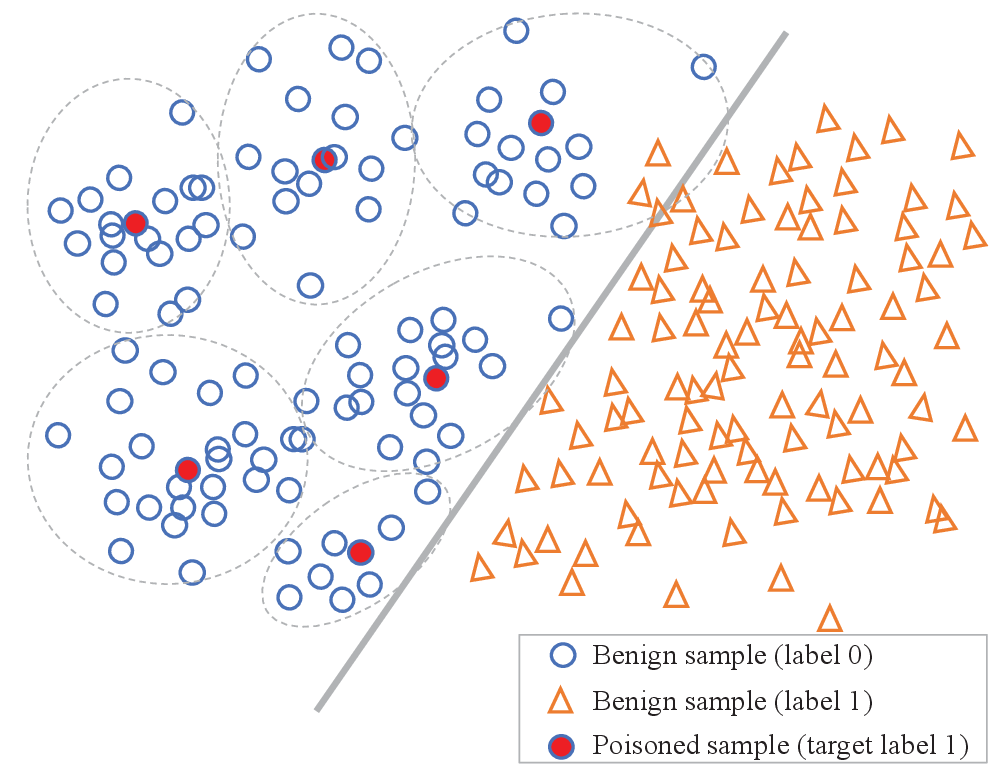}}
\\
\subfloat[]{\label{fig:fig2_c} \includegraphics[width=.4\linewidth,clip]{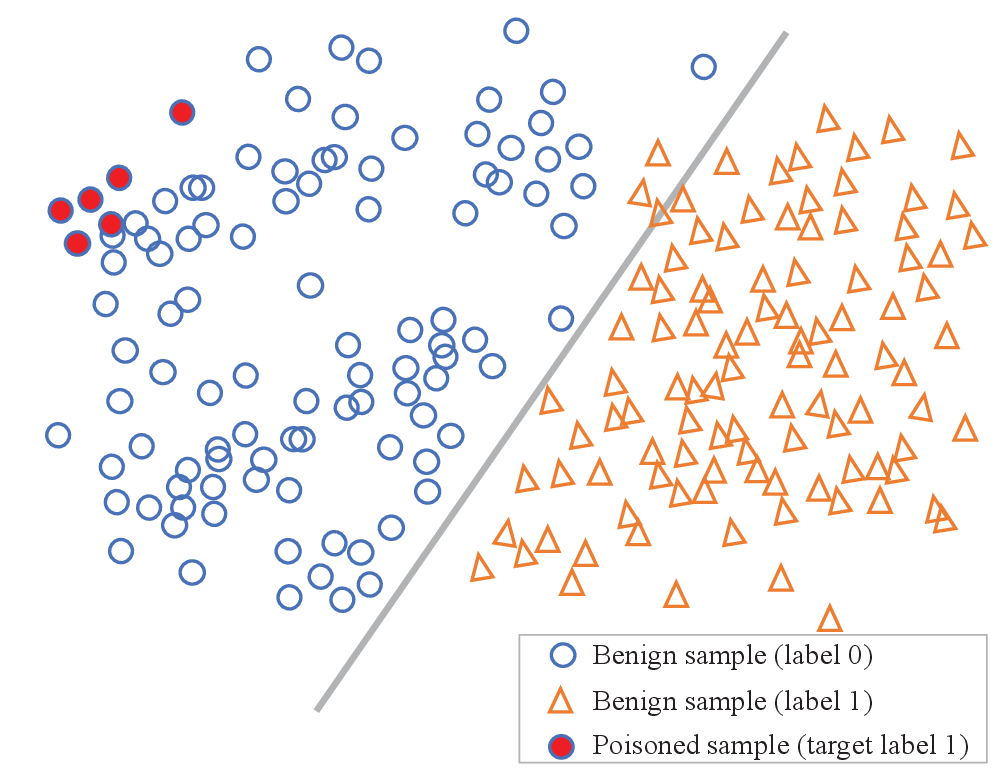}}
\quad
\subfloat[]{\label{fig:fig2_d} \includegraphics[width=.4\linewidth,clip]{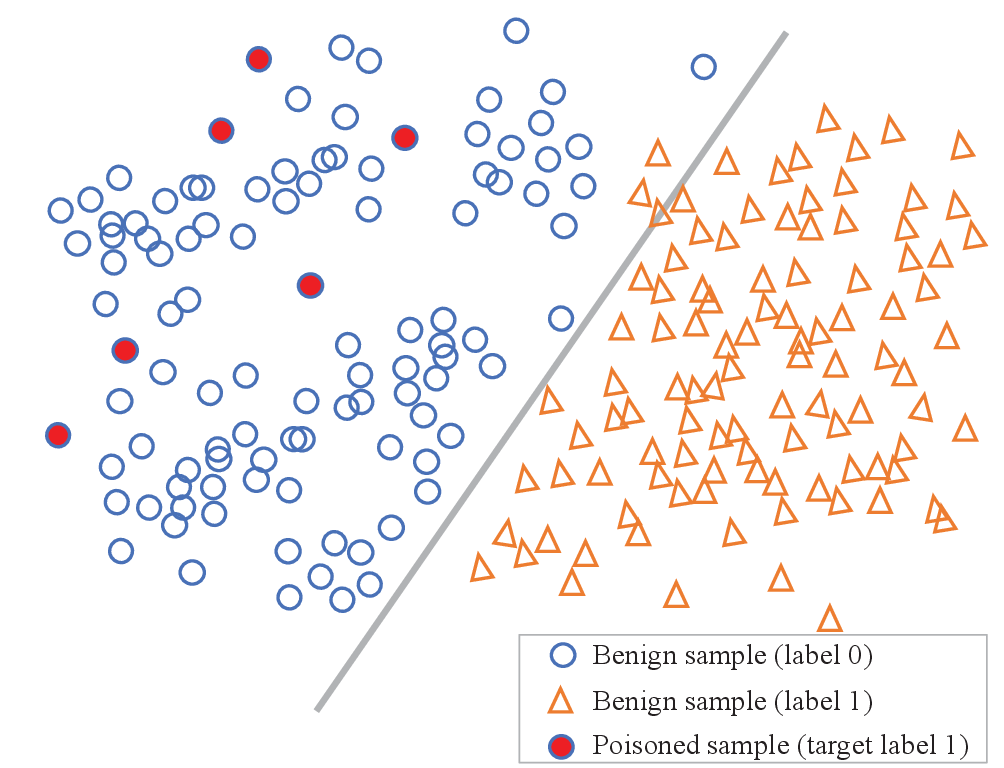}}
\caption{Illustrative examples for AP strategies in binary linear classifier. a) MDS; b) RDS; c) Model-based AP: MUS/MMCS; d) Combinational AP: MUS/MMCS + MDS.} \label{fig:fig2}
\end{figure}

We use the Euclidean distance to measure the diversity. First, we calculate the distance between each sample and other samples in all source EEG data $\mathscr{D}=\{X^n\}_{n=1}^{\sum_{s=1}^{S}N_s}$ and compute
\begin{eqnarray}
d(X^n)=\mathop{\min}_{X^{n'}\in{\mathscr{D}-X^n}} \left\| X^n-X^{n'}\right\|,  \label{eq:MDS}
\end{eqnarray}
i.e., the minimum Euclidean distance between $X^n$ and all other source-domain samples $X^{n'}\in{\mathscr{D}-X^n}$. The top $P$ samples with the maximum diversities are then selected as the candidates for poisoning, as shown in Figure~\ref{fig:fig2_a}.

In addition, the diversity criterion can be combined with a model-based approach to enhance the AP performance, as shown later in this paper.

\subsubsection{Representativeness and diversity sampling (RDS)}

Representativeness is also an important criterion in AL~\cite{drwuSAL2019}. It selects for labeling the samples that can represent as many surrounding samples as possible, i.e., those close to the cluster centers, to avoid selecting outliers. Usually the representativeness and the diversity cannot be maximized simultaneously, so we need to find a good compromise between them.

RDS AP uses a clustering-based approach to select both representative and diverse samples for poisoning. Specifically, we first perform $k$-means ($k=P$) clustering on all source domain EEG samples, and then select $P$ samples closest to the $P$ cluster centers (one sample from each cluster) to poison, as shown in Figure~\ref{fig:fig2_b}. The samples near the cluster centers are most representative among clusters; different clusters scatter the entire feature space, which ensures diversity.

In fact, RDS AP is identical to the RD AL approach proposed in our previous work~\cite{drwuSAL2019}, though the latter was for regression problems.

\subsubsection{Minimum uncertainty sampling (MUS)}

Uncertainty sampling is a classical AL approach, which selects the samples whose estimated labels are most uncertain, e.g., closest to the decision boundary, for labeling. For a classifier with probabilistic output, the uncertainty of an unlabeled sample $\mathbf{x}$ can be measured by the Shannon entropy:
\begin{eqnarray}
    u(\mathbf{x})= -\sum_{y}p(y|\mathbf{x}) \log p(y|\mathbf{x}). \label{eq:entropy}
\end{eqnarray}

In contrast to AL, AP selects the samples with the lowest uncertainty for poisoning in backdoor attacks. The rationale is as follows: since the original model has low uncertainty (high classification confidence) on these samples, e.g., these samples are far away from the classification boundary, as shown in Figure~\ref{fig:fig2_c}, poisoning these samples and changing their labels would bring large distortion to the decision boundary; if the high-confident samples can be misclassified when the trigger is added, then those less-confident ones will also likely be misclassified with the trigger.

To perform MUS, we first train a probabilistic classifier on the EEG samples $\{X^n\}_{n=1}^{\sum_{s=1}^{S}N_s}$ from the $S$ source subjects to obtain the posterior probability of non-target samples, and then calculate the entropy-based uncertainty $u(X^n)$ in (\ref{eq:entropy}). The top $P$ samples with the minimum entropy values are selected as the candidate samples for poisoning. For binary classification, as frequently used in BCIs, this is equivalent to choosing the non-target samples with the largest $p(0|\mathbf{x})$ for poisoning.

\subsubsection{Minimum model change sampling (MMCS)}

Another popular AL approach, expected model change maximization~\cite{cai2017active}, estimates the model parameter change caused by the addition of unlabeled sample(s) and selects those with the maximum change to label. The rationale is that a sample that can bring a big change to the model is usually informative.

MMCS is a model-specific AL approach. When the cross-entropy loss is used in gradient descent optimization of a binary logistic regression classifier $f$, the model change on a sample $\mathbf{x}$ can be approximated as~\cite{cai2017active}
\begin{eqnarray}
c(\mathbf{x})=(f(\mathbf{x})-y)\mathbf{x}, \label{eq:MMC}
\end{eqnarray}
where $y$ is the ground-truth label of the unlabeled sample $\mathbf{x}$, which is unknown in AL but can be estimated using the expectation over all classes.

In contrast to AL, AP chooses the samples that minimize the model change for poisoning in backdoor attacks: if the samples that have the least influence on the model parameters can successfully create a backdoor after poisoning, i.e., the infected model is very sensitive to the backdoor trigger, then other samples that have more influence to the model parameters are also likely to perform backdoor attacks successfully.

More specifically, we first train a source model $f$ on all EEG samples $\{X^n\}_{n=1}^{\sum_{s=1}^{S}N_s}$ to obtain the softmax output of the non-target samples, and use $y=0$ in (\ref{eq:MMC}) to compute the model change of each sample, $c(X^n)$. Finally, we choose the top $P$ samples with the minimum model changes as the candidates for poisoning, as shown in Figure~\ref{fig:fig2_c}.

\subsubsection{Combinational AP (MUS/MMCS+MDS)}

As stated above, MUS and MMCS are model-based AP approaches, which need to train a source model $f$. Their main idea is to select the samples far away from the current decision boundary, as shown in Figure~\ref{fig:fig2_c}, though their implementations are different. The selected samples may be concentrated together, making the embedded backdoor pattern inapplicable to the entire input space.

The diversity-based MDS AP can be combined with the model-based MUS or MMCS AP to select samples for poisoning, by exploiting the input and output information simultaneously, as shown in Figure~\ref{fig:fig2_d}.

Specifically, MUS+MDS measures the uncertainty $-u(X^n)$ in (\ref{eq:entropy}) and the distance $d(X^n)$ in (\ref{eq:MDS}), normalizes their values to the same range, and then adds them up:
\begin{eqnarray}
du(X^n)={\rm Normalize}(d(X^n))-{\rm Normalize}(u(X^n)).
\end{eqnarray}

Similarly, MMCS+MDS computes
\begin{eqnarray}
dc(X^n)={\rm Normalize}(d(X^n))- {\rm Normalize}(c(X^n)).
\end{eqnarray}
Then, we select the top $P$ samples with the maximum $du(X^n)$ in MUS+MDS or $dc(X^n)$ in MMCS+MDS as the candidates for poisoning.

\section{Experiments and results} \label{sect:Experiment}

Experiments were performed in this section to verify the effectiveness of our proposed AP approaches.

\subsection{Datasets}

The following four datasets were used in this study, as in~\cite{zhang2019vulnerability,he2019transfer}, including two datasets (P300 and ERN) on event-related potentials and two MI datasets in BCI Competition IV\footnote{\url{http://www.bbci.de/competition/iv/}}.

\textbf{P300 evoked potentials (P300)}: The P300 dataset~\cite{EPFLP300} was collected from eight subjects, who faced a laptop on which six images were flashed randomly to elicit P300 responses. The goal was to classify whether the image is target or non-target. The EEG data were recorded from 32 channels at 2,048 Hz. We applied a [1,40] Hz band-pass filter and downsampled them to 128 Hz. Next, we extracted the [0,1]s EEG epochs after each image onset and standardized them using $z$-score normalization. Each subject had about 3,300 EEG epochs.

\textbf{Feedback error-related negativity (ERN)}: The ERN dataset~\cite{ERN} was used in a Kaggle competition\footnote{\url{https://www.kaggle.com/c/inria-bci-challenge}} for detecting errors during the P300 spelling task, in order to determine whether the selected item was correct by analyzing the EEG signals after the subject received feedback. It is a binary classification dataset collected from 26 subjects with 56 channels, and partitioned into a training set of 16 subjects and a test set of 10 subjects. Our experiments only used the 16 subjects in the training set. We also applied a [1, 40] Hz band-pass filter, downsampled the EEG signals to 128 Hz, extracted EEG epochs between [0, 1.3]s, and standardized them using $z$-score normalization. Each subject had 340 EEG epochs.

\textbf{MI1}: The MI1 dataset was Dataset 2a\footnote{\url{https://www.bbci.de/competition/iv/desc_2a.pdf}}~\cite{MI4C} in BCI Competition IV. It was collected from nine subjects and included four classes: imagined movements of the left hand, right hand, both feet, and tongue. The 22-channel EEG signals were recorded at 250 Hz. We applied a [8, 30] Hz band-pass filter, downsampled the EEG signals to 128 Hz, extracted EEG epochs between [0.5, 3.5]s, and standardized them using $z$-score normalization. Only two classes (left hand and right hand) were used in our experiments. Each subject had 144 epochs, 72 in each class.

\textbf{MI2}: The MI2 dataset was Dataset 1\footnote{\url{https://www.bbci.de/competition/iv/desc_1.html}}~\cite{blankertz2007non} in BCI Competition IV. It includes 59-channel EEG signals from seven subjects. The same two classes (left hand and right hand) were selected and the same preprocessing as MI1 was carried out. Each subject had 100 epochs per class in the calibration phase with complete marker information.

\subsection{CNN models}

Three CNN models were used in our experiments, as in~\cite{zhang2019vulnerability,drwuUAP2021}:

\textbf{EEGNet}: EEGNet~\cite{EEGNet} has a compact EEG-specific CNN architecture. It contains a temporal convolutional block, a depthwise separable convolutional block and a classification block. The depthwise separable convolution can reduce the number of model parameters and learn a good feature map.

\textbf{DeepCNN}: DeepCNN~\cite{MNE} consists of four convolutional blocks and a classification block. The first convolutional block is a combination of a temporal convolution and a spatial filter for handling multi-channel EEG signals, and the other three are standard convolutional-max-pooling blocks.

\textbf{ShallowCNN}: ShallowCNN~\cite{MNE} is a shallow version of DeepCNN. It has only one convolutional block which is similar to the first convolutional block in DeepCNN, but a larger kernel size, a different activation function, and a different pooling approach.

We used Adam optimizer, cross-entropy loss, and batch size 64. Early stopping was used to reduce over-fitting. For MMCS, we estimated the model change from the change of parameters in the last classification block. MDS and RDS used the input to the last classification block as features.

\subsection{Experimental settings}
\subsubsection{Attack settings}

To simulate cross-subject backdoor attacks in TL-based BCIs, for each dataset, we sequentially selected one subject as the target subject and all remaining ones as the source subjects, i.e., we performed leave-one-subject-out cross-validation to evaluate the attack and classification performance. This procedure was repeated for each subject, so that each one became the target subject once for testing, the remaining ones became the source subjects and were combined for training. For P300 and ERN, we performed under-sampling for each source subject to overcome class imbalance. The source data were further randomly partitioned into 80\% training and 20\% validation for early stopping. The entire leave-one-subject-out cross-validation process was repeated five times, and the mean results are reported in all following subsections.

In each repeat, a small amount of EEG samples in the combined source-domain samples were poisoned randomly (baseline) or by AP, then data from different subjects were aligned by a TL approach in Section~\ref{sec:EA}, and combined to train deep learning models.

The target label for poisoned samples is `target' in P300, `good-feedback' in ERN, and `right hand' in MI1 and MI2. The attacker's goal was to make the model classify any sample in the target domain with the trigger into `target'/`good-feedback'/`right hand'/`right hand' for P300/ERN/MI1/MI2, no matter what true labels they have.

\subsubsection{Performance measures} \label{sec:measures}

We used the following two metrics to evaluate the classification and attack performance:
\begin{enumerate}
\item \emph{Balanced classification accuracy (BCA)}, which is the average of the per-class classification accuracies.
\item \emph{Attack success rate (ASR)}, which is
\begin{eqnarray}
ASR=\frac{\mbox{Number of successfully attacked samples}}{\mbox{Total number of attacked samples}}. \label{eq:ASR}
\end{eqnarray}
A successfully attacked sample is the one with the trigger added and successfully classified into the target class specified by the attacker. To ensure the ASR is caused solely by a backdoor attack instead of inaccurate predictions, we computed the ASR on the correctly classified non-target samples only, i.e., if a non-target sample is already misclassified into the target class without adding the trigger, then it is not counted in neither the numerator nor the denominator of (\ref{eq:ASR}).
\end{enumerate}

The clean target-domain data were used to compute the BCA, and the poisoned target-domain data were used to compute the ASR.

\subsubsection{TL approach} \label{sec:EA}

Since EA has demonstrated outstanding TL performance in both traditional machine learning and deep learning~\cite{he2019transfer,kostas2020thinker}, it was used in our experiments.

For each source domain $\{X_s^n\}_{n=1}^{N_s}$, EA first computes the mean covariance matrix of all EEG trials by
\begin{eqnarray}
\bar{R}_s=\frac{1}{N_s}\sum_{n=1}^{N_s}X_s^n(X_s^n)^\intercal,
\end{eqnarray}
then performs the following transformation to each trial:
\begin{eqnarray}
\hat{X}_s^n=\bar{R}_s^{-1/2}X_s^n.
\end{eqnarray}

$\hat{X}_s^n$ then replaces the original $X_s^n$ in all subsequent signal processing and machine learning.

Similarly, for the target-domain samples, we computed the mean covariance matrix $\bar{R}_t$ and obtained the aligned EEG samples $\hat{X}_t^n$.

In backdoor attacks, we assume the attacker first adds the backdoor to some selected source domain non-target samples, then an innocent user downloads the poisoned dataset and uses it with his/her own data for TL. Thus, EA should be performed after source-domain data poisoning.

\begin{figure*}[htpb] \centering
\includegraphics[width=\linewidth,clip]{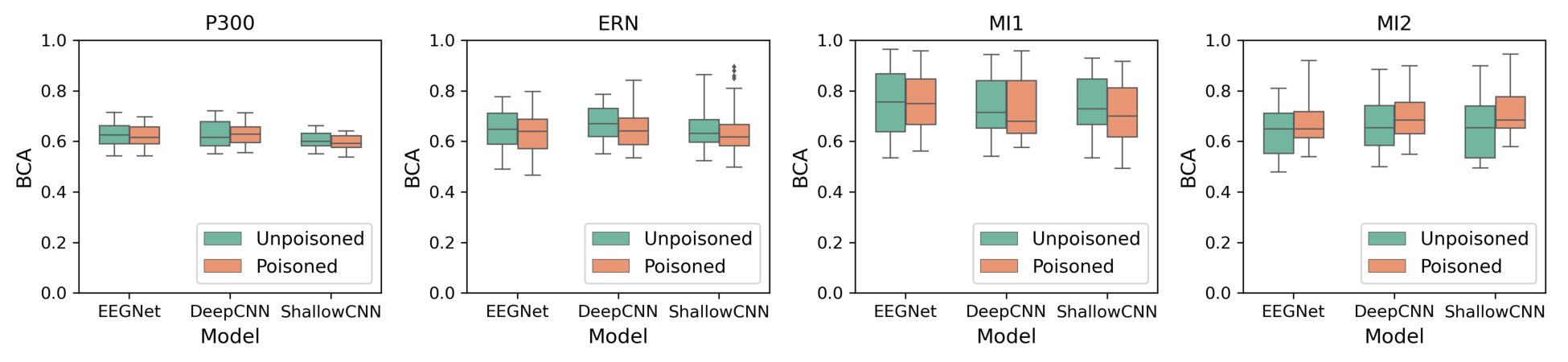}
\caption{BCAs with and without poisoned samples.} \label{fig:fig3}
\end{figure*}

We first compared the BCAs of EA when 5\% source-domain data were poisoned, with those when no data were poisoned at all. The results are shown in Figure~\ref{fig:fig3}. The BCAs with and without data poisoning were very similar, i.e., the injected NPP trigger did not significantly change the TL performance, which is desirable. However, we still want the number of poisoned samples to be as small as possible, since the trigger has a fixed pattern, and taking a simple average of a large amount of poisoned samples may expose it.

\subsection{Attack performance}

\subsubsection{Baseline}

First, we evaluated the baseline performance on the benign models, i.e., the models trained on the clean aligned source-domain data. The BCAs and ASRs of different models and datasets on the target-domain data are shown in the `Baseline' panel of Table~\ref{tab:attack}. The baseline BCAs were obtained without using any labeled target-domain data. They were well above the 50\% chance level for binary classification, indicating the effectiveness of TL. However, the ASRs were mostly close to zero, i.e., when the source-domain data were not deliberately poisoned to embed the backdoor, the trigger had little effect on attacking the target model.

\begin{table*}[htbp]  \center
\caption{Classification and attack performances (\%) with poisoning rate 5\%. The best two ASRs for each model and each dataset are marked in bold.} \setlength{\tabcolsep}{0.8mm} \label{tab:attack} \fontsize{8pt}{\baselineskip}\selectfont
\begin{tabular}{c|c|cc|cc|cc|cc|cc|cc|cc|cc}
\toprule
\multirow{3}{*}{Dataset} &\multirow{3}{*}{Model} &\multicolumn{2}{c}{Baseline}  &\multicolumn{14}{|c}{Active Poisoning}\\ \cline{3-18}
 & &\multirow{2}{*}{BCA} &\multirow{2}{*}{ASR} &\multicolumn{2}{c}{Random} &\multicolumn{2}{|c}{MDS} &\multicolumn{2}{|c}{RDS} &\multicolumn{2}{|c}{MUS} &\multicolumn{2}{|c}{MMCS} &\multicolumn{2}{|c}{MUS+MDS} &\multicolumn{2}{|c}{MMCS+MDS}\\ \cline{5-18}
 &                &                &                &BCA     &ASR       &BCA     &ASR       &BCA     &ASR       &BCA     &ASR       &BCA     &ASR       &BCA     &ASR       &BCA     &ASR \\
\midrule
\multirow{3}{*}{P300}&EEGNet     &$62.5$ &$0.4$ &$62.0$ &$83.2$ &$62.1$ &$80.9$ &$62.6$ &$86.3$ &$61.8$ &$92.1$ &$61.4$ &$\mathbf{94.3}$ &$62.1$ &$92.0$ &$62.3$ &$\mathbf{93.2}$\\
                     &DeepCNN    &$62.9$ &$0.1$ &$62.9$ &$79.0$ &$63.2$ &$82.3$ &$63.5$ &$84.9$ &$63.0$ &$88.8$ &$63.1$ &$91.3$ &$62.8$ &$\mathbf{91.6}$ &$63.3$ &$\mathbf{92.9}$\\
                     &ShallowCNN &$60.6$ &$0.3$ &$59.7$ &$31.8$ &$60.0$ &$41.6$ &$60.3$ &$43.0$ &$59.6$ &$40.2$ &$59.3$ &$43.9$ &$60.3$ &$\mathbf{47.3}$ &$59.6$ &$\mathbf{50.4}$\\
                     \midrule
\multirow{3}{*}{ERN} &EEGNet     &$64.9$ &$1.5$ &$63.7$ &$78.1$ &$63.9$ &$87.6$ &$63.7$ &$87.9$ &$62.9$ &$91.7$ &$63.3$ &$92.7$ &$63.1$ &$\mathbf{93.6}$ &$63.0$ &$\mathbf{94.0}$\\
                     &DeepCNN    &$66.6$ &$2.4$ &$65.0$ &$75.8$ &$64.7$ &$86.9$ &$65.8$ &$87.9$ &$64.4$ &$90.0$ &$64.8$ &$92.0$ &$64.7$ &$\mathbf{92.3}$ &$64.9$ &$\mathbf{92.4}$\\
                     &ShallowCNN &$64.7$ &$7.1$ &$63.8$ &$38.1$ &$63.2$ &$70.4$ &$63.4$ &$72.5$ &$62.6$ &$71.4$ &$62.8$ &$71.2$ &$62.7$ &$\mathbf{77.5}$ &$62.7$ &$\mathbf{77.1}$\\
                     \midrule
\multirow{3}{*}{MI1}  &EEGNet     &$75.7$ &$4.5$ &$75.7$ &$96.6$ &$76.2$ &$99.0$ &$76.6$ &$98.2$ &$75.6$ &$\mathbf{99.3}$ &$75.9$ &$99.1$ &$76.7$ &$\mathbf{99.2}$ &$76.5$ &$98.9$\\
                     &DeepCNN    &$74.1$ &$0.3$ &$72.7$ &$61.6$ &$74.4$ &$82.3$ &$74.0$ &$86.2$ &$72.7$ &$86.2$ &$73.7$ &$\mathbf{87.4}$ &$73.3$ &$\mathbf{86.7}$ &$74.0$ &$85.2$\\
                     &ShallowCNN &$74.0$ &$0.0$ &$71.1$ &$5.0$  &$70.7$ &$17.4$ &$71.1$ &$24.3$ &$67.9$ &$14.5$ &$68.6$ &$17.0$ &$70.0$ &$\mathbf{30.3}$ &$70.6$ &$\mathbf{29.9}$\\
                     \midrule
\multirow{3}{*}{MI2}  & EEGNet & 63.8  & 0.7   & 67.1  & 73.4  & 68.2  & 82.4  & 68.4  & 83.5  & 67.2  & 81.7  & 67.5  & 83.1  & 68.4  & \textbf{85.7 } & 68.1  & \textbf{84.8 } \\
          & DeepCNN & 66.9  & 0.2   & 69.9  & 56.1  & 71.3  & 79.5  & 71.4  & \textbf{86.9 } & 68.3  & 81.3  & 69.8  & 83.3  & 70.6  & \textbf{85.4 } & 71.2  & 85.0  \\
          & ShallowCNN & 65.9  & 12.5  & 71.5  & 33.1  & 72.3  & 73.0  & 70.0  & 91.9  & 70.4  & 58.4  & 70.5  & 63.3  & 71.1  & \textbf{92.6 } & 70.6  & \textbf{92.0 } \\
\midrule
\multirow{3}{*}{Average} & EEGNet & 66.7  & 1.8   & 67.1  & 82.8  & 67.6  & 87.4  & 67.8  & 89.0  & 66.9  & 91.2  & 67.0  & 92.3  & 67.6  & \textbf{92.6 } & 67.5  & \textbf{92.7 } \\
          & DeepCNN & 67.6  & 0.7   & 67.7  & 68.1  & 68.4  & 82.8  & 68.6  & 86.4  & 67.1  & 86.6  & 67.9  & 88.5  & 67.8  & \textbf{89.0 } & 68.4  & \textbf{88.9 } \\
          & ShallowCNN & 66.3  & 5.0   & 66.5  & 27.0  & 66.6  & 50.6  & 66.2  & 57.9  & 65.1  & 46.1  & 65.3  & 48.8  & 66.0  & \textbf{61.9 } & 65.9  & \textbf{62.3 } \\
\bottomrule
\end{tabular}
\end{table*} \normalsize

\subsubsection{AP attacks} \label{sec:ap_attack}

We used the NPP trigger with period $T=1$s, duty cycle $d=20$\% and a random phase $\phi$ in [0, 0.8]$T$ for all poisoned samples. The amplitude $a$ was set to 0.2\%, 15\%, 30\% and 100\% of the mean channel-wise standard deviation of the EEG amplitude for P300, MI1, MI2 and ERN, respectively. Different amplitudes were used on the three datasets because the ranges of EEG amplitudes in different datasets varied significantly.

\begin{figure}[htbp]\centering
\subfloat[]{\label{fig:NPP} \includegraphics[width=.45\linewidth,clip]{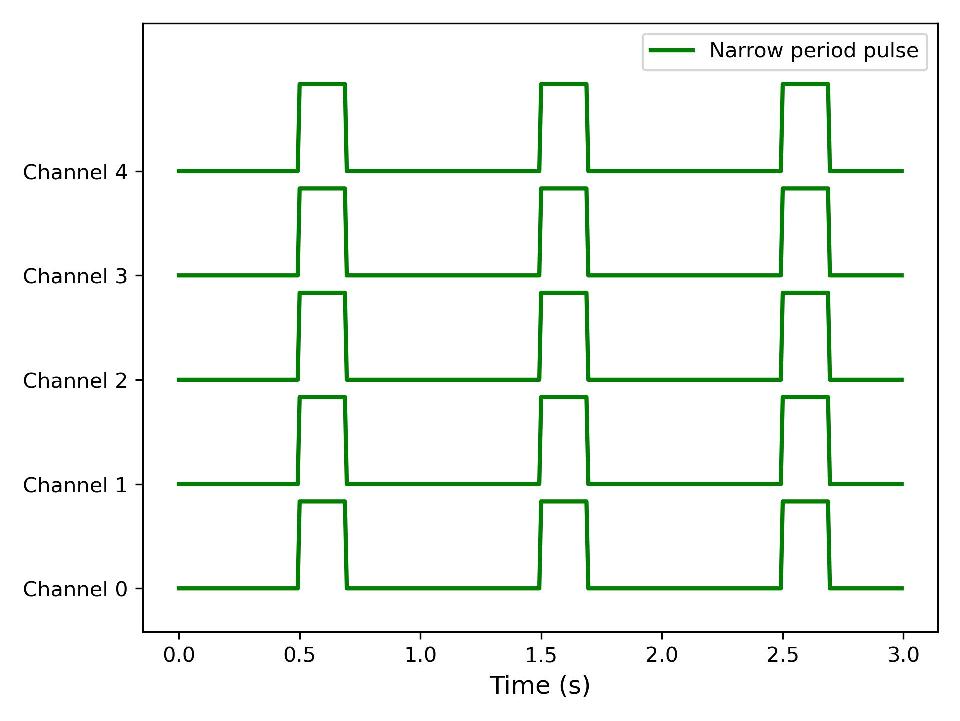}}
\quad \quad
\subfloat[]{\label{fig:EEG} \includegraphics[width=.45\linewidth,clip]{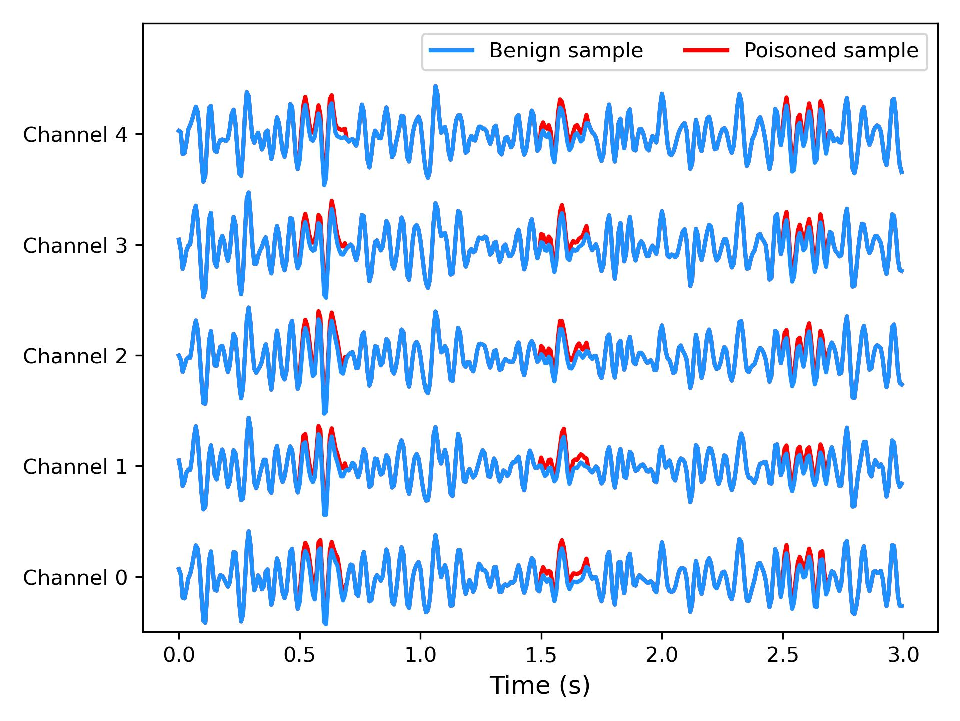}}
\caption{(a) The NPP trigger and (b) EEG signal of the first five channels before and after poisoning. Best viewed in color.} \label{fig:fig4}
\end{figure}

An example of the added NPP trigger and the EEG signal before and after poisoning MI1 is shown in Figure~\ref{fig:fig4}. The poisoned sample with the NPP trigger is almost identical to the benign sample (NPP with $a=30$\%, $d=20$\% and $\phi=0.5T$ was used for this example), making the trigger very difficult to notice.

When the poisoning rate was 5\%, the classification and attack performances of different AP strategies are shown in the `Active Poisoning' panel of Table~\ref{tab:attack}, where `Random' means the poisoned samples were randomly selected from the non-target class of the source data, and the others were selected by various AP strategies from the non-target class of the source data. Table~\ref{tab:attack} shows that:

\begin{enumerate}
\item The BCAs of all AP approaches were very close to those in Baseline, indicating that introducing poisoned samples or embedding a backdoor did not significantly degrade the normal classification performance, if the input target-domain samples did not contain the trigger.
\item The ASRs of different AP approaches were significantly improved over Baseline, indicating the effectiveness of backdoor attacks, i.e., once the NPP trigger was added to a non-target sample from the target subject, the model would very likely misclassify it into the target class.
\item Our proposed AP strategies, including MDS, RDS, MUS, MMCS and their combinations, generally achieved higher ASRs on different models and datasets, compared with Random, e.g., 83.2\% (Random) versus 94.3\% (MMCS) for EEGNet on P300, indicating that AP can improve the attack efficiency under the same poisoning rate.
\item The ASRs of MUS and MMCS were generally higher than those of MDS and RDS, likely because MUS and MMCS are supervised and model-based approaches, which can utilize more information than the unsupervised MDS and RDS approaches.
\item The attack performance of RDS was better than that of MDS, suggesting that it is better to consider both representativeness and diversity in AP than diversity only.
\item The ASRs of the combinational approaches that integrate MUS/MMCS with MDS (the last two columns in Table~\ref{tab:attack}) were generally higher than those of MUS/MMCS, indicating that considering uncertainty/model change and diversity simultaneously helped improve the attack efficiency.
\end{enumerate}

\begin{figure}[htbp]\centering
\subfloat[]{\label{fig:fig5_AL_P300}   \includegraphics[width=\linewidth, trim=100 0 100 0, clip]{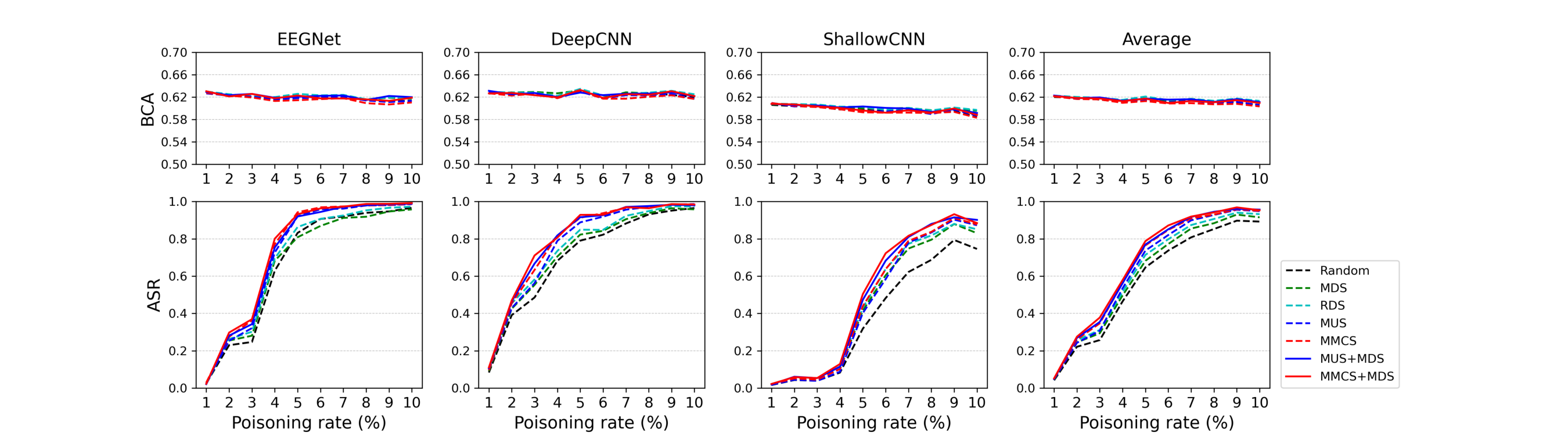}}
\quad
\subfloat[] {\label{fig:fig5_AL_ERN}   \includegraphics[width=\linewidth, trim=100 0 100 0,clip]{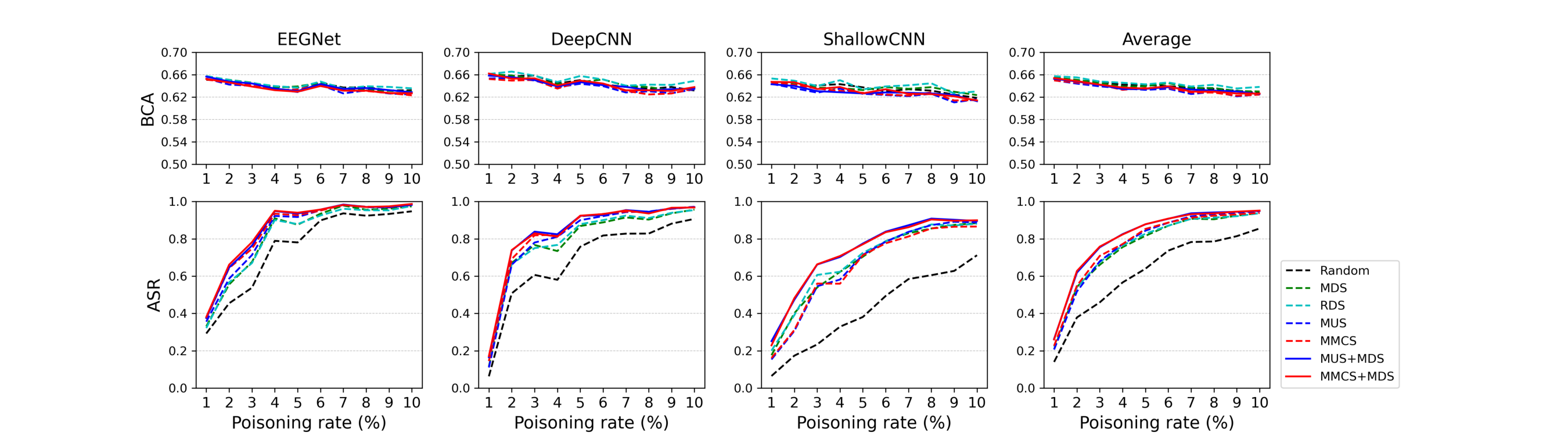}}
\quad
\subfloat[] {\label{fig:fig5_AL_MI1}   \includegraphics[width=\linewidth, trim=100 0 100 0,clip]{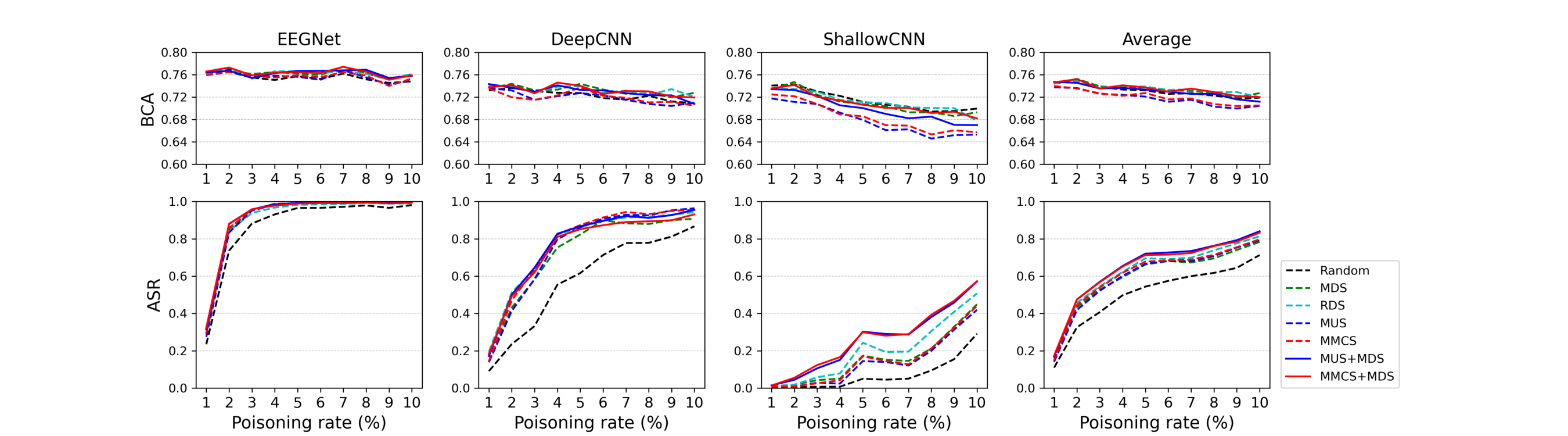}}
\quad
\subfloat[] {\label{fig:fig5_AL_MI2}   \includegraphics[width=\linewidth, trim=100 0 100 0,clip]{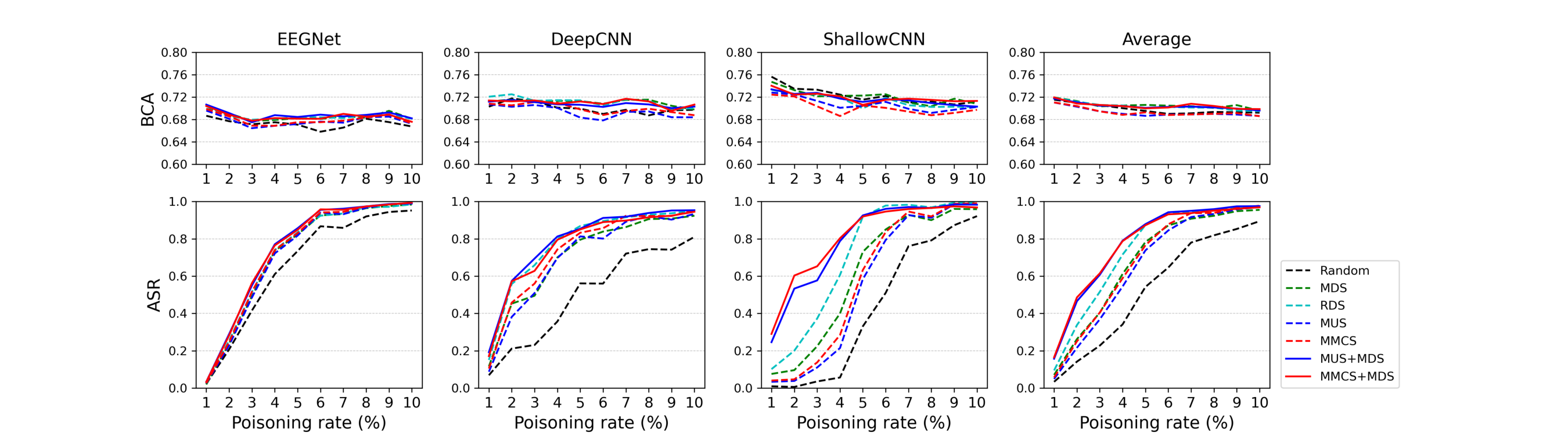}}
\caption{BCAs and ASRs at different poisoning rates on a) P300; b) ERN; c) MI1; and, d) MI2.} \label{fig:fig5_AL}
\end{figure}

Figure~\ref{fig:fig5_AL} shows the BCAs and ASRs at different poisoning rates on the four datasets. The parameters of the NPP trigger on each dataset were the same for different deep learning models, so the robustness of different models can be compared. Observe that:
\begin{enumerate}
\item Generally, as the poisoning rate increased from 1\% to 10\%, the BCAs of all AP strategies remained stable and comparable to those of Baseline, indicating that backdoor attacks in TL did not degrade the classification performance on normal samples and was difficult to detect.
\item The ASRs of all AP strategies on all four datasets and for all three deep learning models increased rapidly as the poisoning rate increased, especially for EEGNet and DeepCNN. The ASRs of ShallowCNN  were relatively low when the poisoning rate was small, likely because ShallowCNN has small capacity to remember the trigger pattern, which was also found in~\cite{yao2019latent,wang2020attack}.
\item Generally, our proposed AP strategies achieved higher ASRs than Random. As the poisoning rate increased, the ASR improvement of AP gradually vanished. This is consistent with traditional AL approaches. As more poisoned samples make the backdoor easier to detect, we prefer a small poisoning rate in practice, and hence the proposed AP approaches are desirable.
\item Consistent with the observations from Table~\ref{tab:attack}, generally the combinational AP strategies achieved higher ASRs.
\end{enumerate}

\subsection{Model consistency}

As demonstrated in Section~\ref{sec:AP}, MUS and MMCS are model-based approaches, i.e., they select the most useful samples for poisoning based on the predictions of the model trained on the source data. In previous experiments, the attacker model used in these AP approaches was consistent with the user model in TL, e.g., when the user model in TL was EEGNet, the AP approaches also used EEGNet. However, in practice the attacker does not know which machine learning model the user would use.

\begin{figure}[htbp]\centering
\subfloat[]{\includegraphics[width=\linewidth,clip]{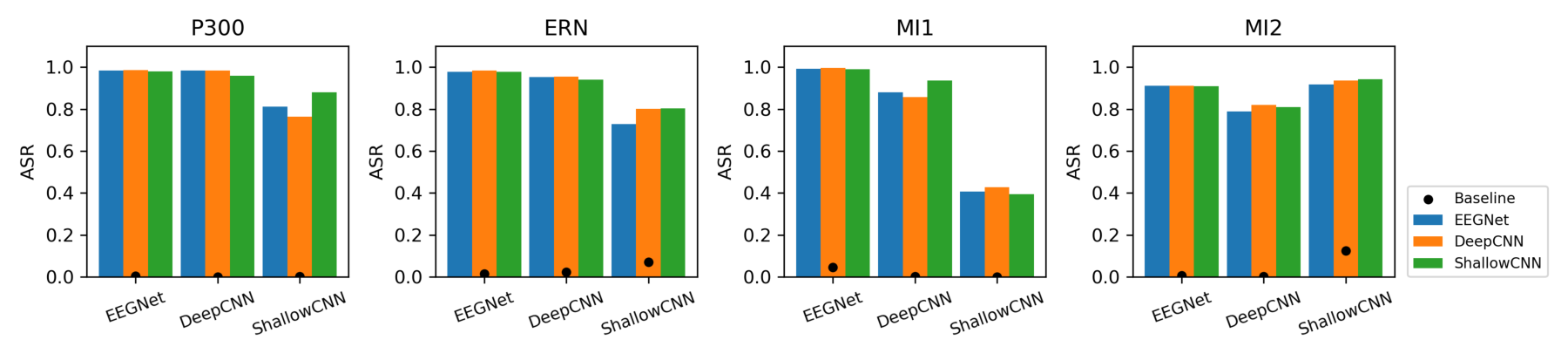}}
\quad
\subfloat[]{\includegraphics[width=\linewidth,clip]{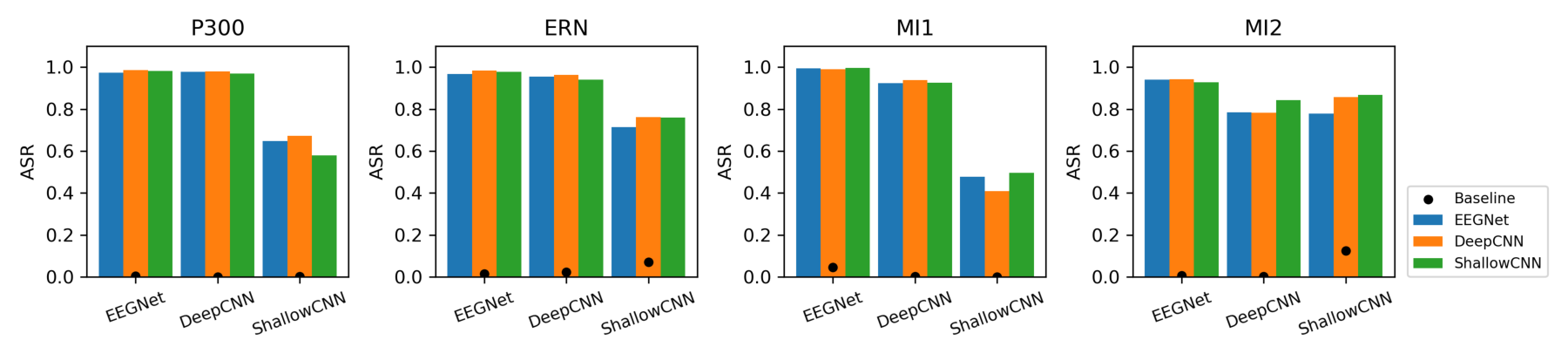}}
\caption{ASRs when the attacker and the user use different models. a) MUS; b) MMCS.} \label{fig:fig6}
\end{figure}

This subsection studies how the model consistency affects the attack performance. The ASRs when MUS and MMCS used different models from the user model in TL, with 10\% poisoning rate, are shown in Figure~\ref{fig:fig6}. The horizontal axis represents the user model, and different bars represent the ASRs when using different MUS and MMCS models. `Baseline' (black dot) represent the ASRs when the source-domain data were not poisoned at all. Clearly, both MUS and MMCS always achieved much higher ASRs than Baseline, regardless of the user model. That is, although the AP approaches are model-based, they do not require the machine learning model to be consistent with the one used by the user in TL for good attack performance. This makes backdoor attacks much easier to perform in practice, and also more dangerous.

\subsection{Stability of AP}

\begin{figure}[htpb]\centering
\includegraphics[width=.9\linewidth,clip]{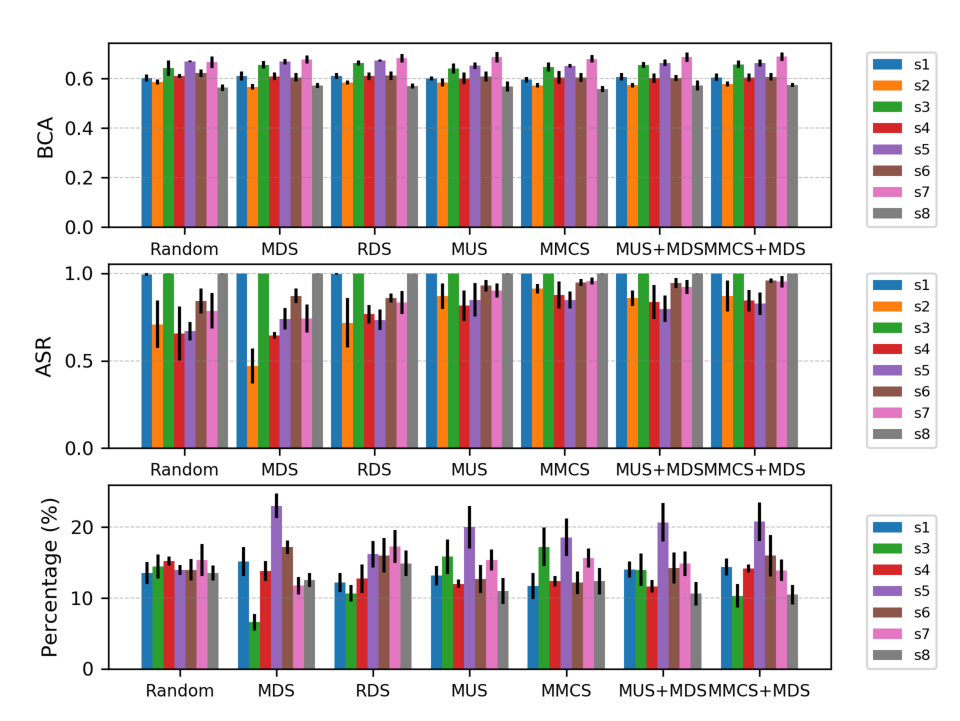}
\caption{Stability analysis using EEGNet.} \label{fig:fig7}
\end{figure}

\begin{figure}[htpb]\centering
\includegraphics[width=.9\linewidth,clip]{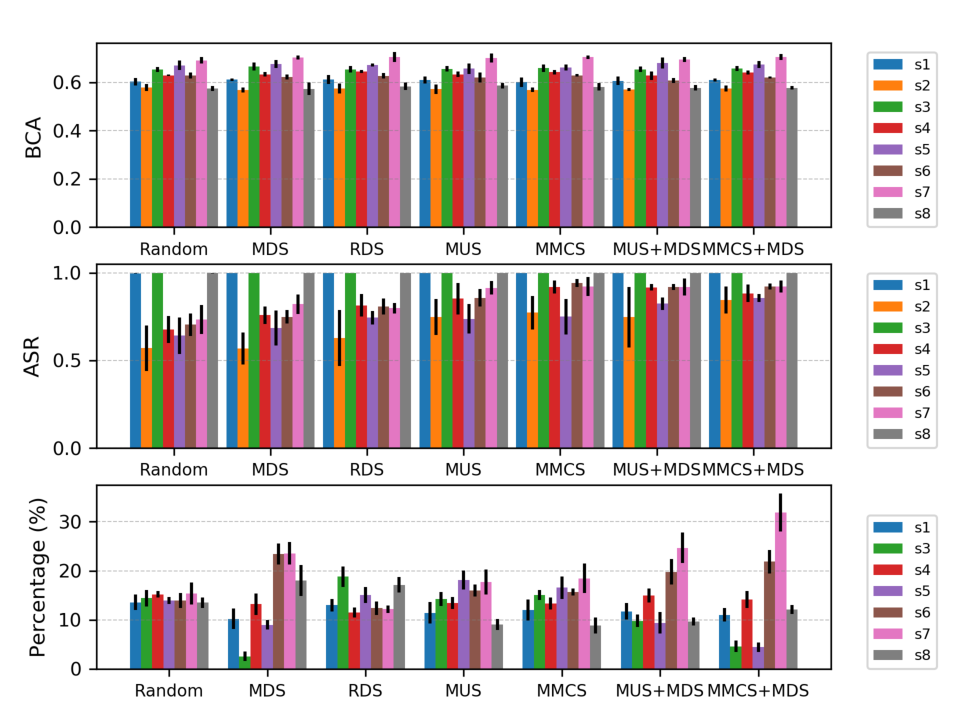}
\caption{Stability analysis using DeepCNN.} \label{fig:fig8}
\end{figure}

\begin{figure}[htpb]\centering
\includegraphics[width=.9\linewidth,clip]{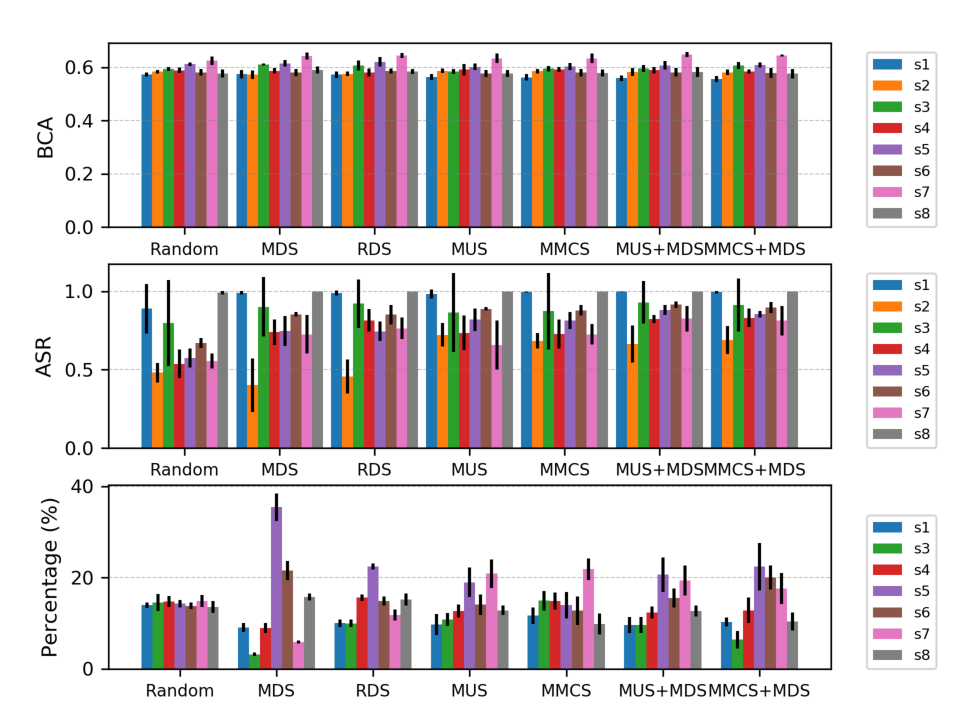}
\caption{Stability analysis using ShallowCNN.} \label{fig:fig9}
\end{figure}

The above experimental results showed the average performance of all subjects on each dataset. In order to demonstrate the stability of our proposed AP approaches across different subjects, we further analyzed the results on each subject on P300 individually, as shown in Figures~\ref{fig:fig7}-\ref{fig:fig9}. We computed the average ASRs from five runs for each subject on P300 using three different deep learning models. The poisoning rates of all AP approaches on EEGNet/DeepCNN/ShallowCNN were 5\%/5\%/8\%, respectively.

Additionally, we computed the number of samples selected for poisoning by different AP approaches from each source subject, when Subject 2 was the target subject (the reason for choosing this subject was that the ASR on this subject was relatively low). The results are shown in the third row of each figure.

Figures~\ref{fig:fig7}-\ref{fig:fig9} show that:
\begin{enumerate}
\item The BCAs of each subject for different deep learning models were similar, all around 0.6. Different AP approaches did not change the BCAs on different subjects.
\item The ASRs of different AP approaches were higher than those of Random. Although there were obvious differences in the ASRs on different subjects due to individual differences, especially Subject 1, Subject 3 and Subject 8, the ASRs almost reached 1 for EEGNet and DeepCNN, and were also very high for ShallowCNN. The ASRs of our proposed AP approaches were generally higher than those of Random for subjects whose attack performance was not very good.
\item The third row of the figures show that the number of samples selected from different source subjects by different AP approaches varied greatly. Random selected about the same number of poisoned samples from each subject, which is intuitive. The numbers of samples selected by MDS and the two model-based strategies (MUS and MMCS) from different subjects were quite different, indicating that not all source subjects were equal in data poisoning. The distributions of the selected samples from different subjects were similar for the two model-based AP approaches, likely because they both used information about the model. It seems that poisoning samples in Subject 3, Subject 5 and Subject 7 was more effective in improving the ASR of Subject 2. Our future work will further investigate backdoor attacks that are robust to individual differences.
\end{enumerate}

\subsection{Influence of trigger} \label{sect:trigger}

\begin{figure}[htbp]\centering
\subfloat[]{\includegraphics[width=.3333\linewidth,clip]{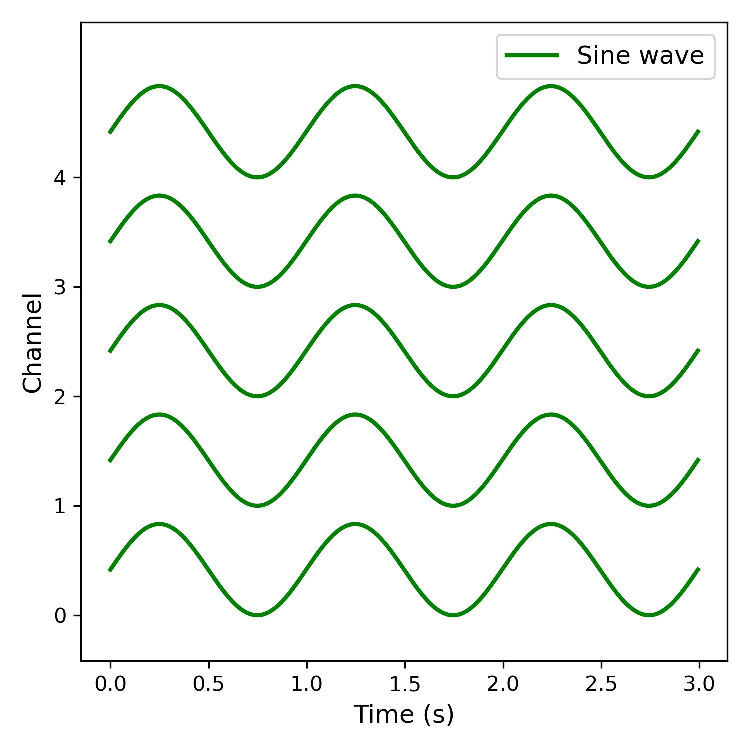}}
\subfloat[]{\includegraphics[width=.3333\linewidth,clip]{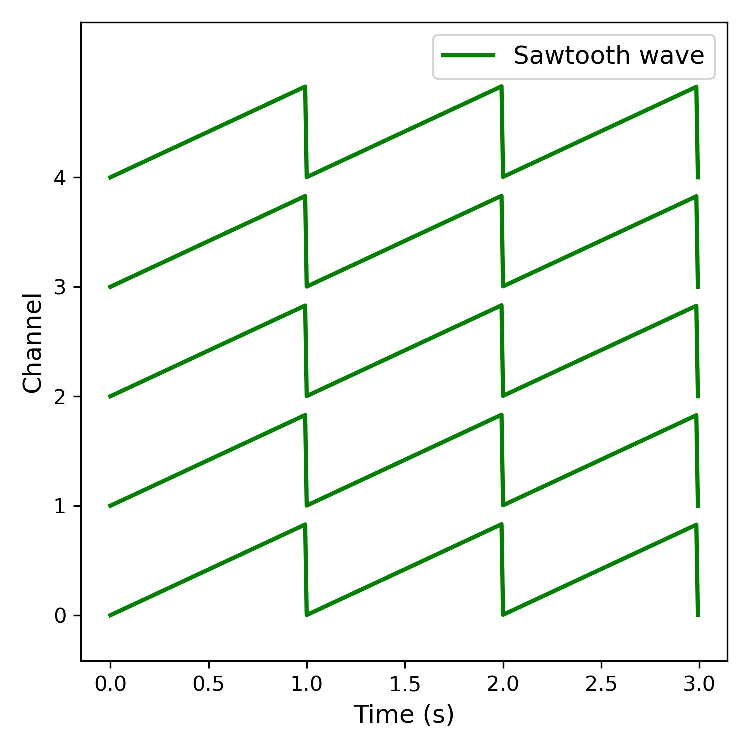}}
\subfloat[]{\includegraphics[width=.3333\linewidth,clip]{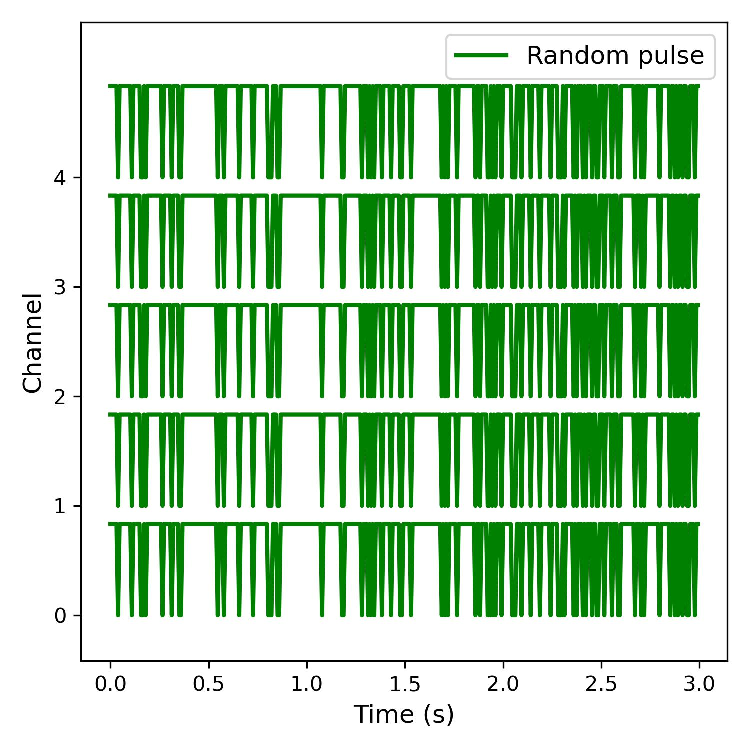}}
\caption{Different triggers on MI1. a) Sine wave; b) Sawtooth wave; c) Random pulse.} \label{fig:triggers}
\end{figure}

We designed three additional triggers as shown in Figure~\ref{fig:triggers} to investigate the influence of trigger on AP approaches. The sine wave and sawtooth wave used the same period, $T=1$s, as NPP trigger. The random pulse was $\mathbf{x}^*=\mathop{\rm sign}(\mathcal{U}(-0.2,0.8))$, where $\mathcal{U}(-0.2,0.8)$ was uniform noise in $[-0.2,0.8]$. Min-max normalization was used to normalize the three types of triggers to [0,1]. Finally, they were multiplied by the same amplitude $a$ to form the triggers.

Tables~\ref{tab:trigger_P300}-\ref{tab:trigger_MI2} show the results of the three triggers on the four datasets. Clearly, backdoor attacks using different triggers were still effective, and the ASRs were generally even higher than those of NPP. In addition, our proposed AP approaches had higher ASRs than Random in most cases, suggesting the robustness of backdoor attacks and AP approaches to different types of triggers.

In practice, we can select a trigger that is easy to generate and reproduce, and robust for backdoor attacks. This study selects the NPP.

\begin{table}[htbp]
  \centering
  \caption{Classification and attack performances (\%) with poisoning rate 5\% and different backdoor triggers on P300. `Average' was calculated excluding `Random'. Average ASRs higher than Random are marked in bold.} \setlength{\tabcolsep}{0.6mm} \fontsize{7pt}{\baselineskip}\selectfont
    \begin{tabular}{c|c|cc|cc|cc|cc|cc|cc|cc|cc}
    \toprule
    \multicolumn{1}{c|}{\multirow{2}[4]{*}{Trigger}} & \multirow{2}[4]{*}{Model} & \multicolumn{2}{c|}{Random} & \multicolumn{2}{c|}{MDS} & \multicolumn{2}{c|}{RDS} & \multicolumn{2}{c|}{MUS} & \multicolumn{2}{c|}{MMCS} & \multicolumn{2}{c|}{MUS+MDS} & \multicolumn{2}{c|}{MMCS+MDS} & \multicolumn{2}{c}{Average} \\
\cline{3-18}          &       & BCA   & ASR   & BCA   & ASR   & BCA   & ASR   & BCA   & ASR   & BCA   & ASR   & BCA   & ASR   & BCA   & ASR   & BCA   & ASR \\
    \midrule
    \multirow{3}[2]{*}{Sine wave} & EEGNet & 62.0  & 94.8  & 62.3  & 94.2  & 62.3  & 96.1  & 61.8  & 98.3  & 61.8  & 98.8  & 62.3  & 98.6  & 62.1  & 98.8  & 62.1  & \textbf{97.5 } \\
          & DeepCNN & 62.4  & 90.4  & 63.4  & 93.3  & 62.6  & 91.7  & 62.2  & 96.4  & 63.0  & 97.8  & 62.5  & 97.5  & 62.8  & 97.1  & 62.8  & \textbf{95.6 } \\
          & ShallowCNN & 60.3  & 65.3  & 60.5  & 78.8  & 60.6  & 78.5  & 60.0  & 76.5  & 60.2  & 82.6  & 60.0  & 85.1  & 60.4  & 86.0  & 60.3  & \textbf{81.3 } \\
    \midrule
    \multirow{3}[2]{*}{Sawtooth wave} & EEGNet & 62.3  & 95.7  & 61.8  & 95.6  & 62.1  & 97.2  & 62.2  & 98.5  & 61.7  & 99.0  & 62.0  & 98.6  & 62.3  & 99.2  & 62.0  & \textbf{98.0 } \\
          & DeepCNN & 62.5  & 86.4  & 62.7  & 89.8  & 62.4  & 90.5  & 62.1  & 96.2  & 62.6  & 97.3  & 62.5  & 96.8  & 62.3  & 96.3  & 62.4  & \textbf{94.5 } \\
          & ShallowCNN & 59.7  & 28.4  & 60.2  & 37.7  & 60.2  & 41.3  & 59.6  & 36.2  & 59.5  & 40.3  & 60.1  & 46.6  & 60.0  & 49.8  & 59.9  & \textbf{42.0 } \\
    \midrule
    \multirow{3}[2]{*}{Random pulse} & EEGNet & 62.1  & 97.0  & 62.1  & 97.0  & 62.3  & 98.1  & 62.3  & 98.8  & 62.3  & 99.0  & 62.5  & 99.0  & 62.7  & 99.5  & 62.4  & \textbf{98.6 } \\
          & DeepCNN & 62.9  & 95.6  & 62.7  & 96.7  & 62.5  & 97.4  & 62.7  & 98.7  & 62.8  & 99.0  & 63.2  & 98.7  & 62.7  & 99.0  & 62.8  & \textbf{98.2 } \\
          & ShallowCNN & 60.2  & 74.2  & 60.4  & 82.6  & 60.6  & 85.6  & 60.0  & 85.8  & 59.7  & 88.5  & 59.9  & 89.5  & 60.4  & 90.7  & 60.2  & \textbf{87.1 } \\
    \bottomrule
    \end{tabular}\normalsize
  \label{tab:trigger_P300}%
\end{table}%

\begin{table}[htbp]
  \centering
  \caption{Classification and attack performances (\%) with poisoning rate  5\% and different backdoor triggers on ERN. `Average' was calculated excluding `Random'. Average ASRs higher than Random are marked in bold.} \setlength{\tabcolsep}{0.6mm} \fontsize{7pt}{\baselineskip}\selectfont
    \begin{tabular}{c|c|cc|cc|cc|cc|cc|cc|cc|cc}
    \toprule
    \multicolumn{1}{c|}{\multirow{2}[4]{*}{Trigger}} & \multirow{2}[4]{*}{Model} & \multicolumn{2}{c|}{Random} & \multicolumn{2}{c|}{MDS} & \multicolumn{2}{c|}{RDS} & \multicolumn{2}{c|}{MUS} & \multicolumn{2}{c|}{MMCS} & \multicolumn{2}{c|}{MUS+MDS} & \multicolumn{2}{c|}{MMCS+MDS} & \multicolumn{2}{c}{Average} \\
\cline{3-18}          &       & BCA   & ASR   & BCA   & ASR   & BCA   & ASR   & BCA   & ASR   & BCA   & ASR   & BCA   & ASR   & BCA   & ASR   & BCA   & ASR \\
    \midrule
    \multirow{3}[2]{*}{Sine wave} & EEGNet & 63.5  & 94.5  & 63.8  & 97.9  & 64.0  & 97.1  & 63.3  & 98.2  & 63.6  & 98.4  & 63.7  & 98.8  & 63.7  & 98.8  & 63.7  & \textbf{98.2 } \\
          & DeepCNN & 65.2  & 93.7  & 66.0  & 97.9  & 65.7  & 96.4  & 64.8  & 98.4  & 65.3  & 98.3  & 65.8  & 98.5  & 65.7  & 98.5  & 65.6  & \textbf{98.0 } \\
          & ShallowCNN & 64.1  & 67.9  & 63.9  & 90.6  & 64.7  & 88.7  & 63.4  & 90.1  & 63.4  & 87.9  & 63.9  & 92.2  & 63.8  & 93.0  & 63.8  & \textbf{90.4 } \\
    \midrule
    \multirow{3}[2]{*}{Sawtooth wave} & EEGNet & 64.0  & 97.2  & 64.3  & 98.7  & 64.9  & 98.3  & 64.1  & 98.8  & 64.3  & 98.9  & 64.3  & 99.0  & 64.1  & 99.0  & 64.3  & \textbf{98.8 } \\
          & DeepCNN & 64.9  & 93.6  & 65.0  & 98.0  & 65.2  & 97.6  & 64.5  & 98.1  & 64.4  & 98.3  & 64.5  & 98.6  & 64.3  & 98.5  & 64.6  & \textbf{98.2 } \\
          & ShallowCNN & 63.9  & 65.6  & 64.2  & 89.7  & 64.5  & 88.9  & 63.5  & 87.3  & 63.6  & 85.5  & 63.7  & 92.3  & 63.6  & 92.9  & 63.8  & \textbf{89.4 } \\
    \midrule
    \multirow{3}[2]{*}{Random pulse} & EEGNet & 64.0  & 98.2  & 64.2  & 99.7  & 64.2  & 99.2  & 63.9  & 99.8  & 64.0  & 99.7  & 64.1  & 99.7  & 63.8  & 99.7  & 64.0  & \textbf{99.6 } \\
          & DeepCNN & 65.2  & 96.5  & 65.4  & 99.3  & 65.8  & 98.6  & 65.6  & 99.4  & 65.7  & 99.2  & 65.0  & 99.5  & 64.7  & 99.3  & 65.4  & \textbf{99.2 } \\
          & ShallowCNN & 63.9  & 85.6  & 64.3  & 97.6  & 64.2  & 95.5  & 63.4  & 96.2  & 63.5  & 95.3  & 63.0  & 97.4  & 63.6  & 98.1  & 63.7  & \textbf{96.7 } \\
    \bottomrule
    \end{tabular}\normalsize
  \label{tab:trigger_ERN}%
\end{table}%

\begin{table}[htbp]
  \centering
  \caption{Classification and attack performances (\%) with poisoning rate 5\% and different backdoor triggers on MI1. `Average' was calculated excluding `Random'. Average ASRs higher than Random are marked in bold.} \setlength{\tabcolsep}{0.6mm} \fontsize{7pt}{\baselineskip}\selectfont
    \begin{tabular}{c|c|cc|cc|cc|cc|cc|cc|cc|cc}
    \toprule
    \multicolumn{1}{c|}{\multirow{2}[4]{*}{Trigger}} & \multirow{2}[4]{*}{Model} & \multicolumn{2}{c|}{Random} & \multicolumn{2}{c|}{MDS} & \multicolumn{2}{c|}{RDS} & \multicolumn{2}{c|}{MUS} & \multicolumn{2}{c|}{MMCS} & \multicolumn{2}{c|}{MUS+MDS} & \multicolumn{2}{c|}{MMCS+MDS} & \multicolumn{2}{c}{Average} \\
\cline{3-18}          &       & BCA   & ASR   & BCA   & ASR   & BCA   & ASR   & BCA   & ASR   & BCA   & ASR   & BCA   & ASR   & BCA   & ASR   & BCA   & ASR \\
    \midrule
    \multirow{3}[2]{*}{Sine wave} & EEGNet & 74.1  & 98.3  & 76.0  & 99.4  & 76.5  & 99.2  & 75.6  & 99.6  & 75.8  & 99.4  & 76.4  & 99.6  & 76.2  & 99.6  & 76.1  & \textbf{99.5 } \\
          & DeepCNN & 72.9  & 85.6  & 73.6  & 94.4  & 74.2  & 94.8  & 73.8  & 97.0  & 73.4  & 97.2  & 73.7  & 96.9  & 73.0  & 95.9  & 73.6  & \textbf{96.0 } \\
          & ShallowCNN & 71.8  & 53.3  & 70.7  & 78.6  & 71.0  & 83.1  & 69.8  & 77.8  & 69.8  & 77.1  & 70.4  & 84.2  & 71.2  & 82.4  & 70.5  & \textbf{80.5 } \\
    \midrule
    \multirow{3}[2]{*}{Sawtooth wave} & EEGNet & 74.1  & 97.0  & 75.4  & 99.1  & 75.9  & 98.2  & 75.5  & 99.3  & 75.1  & 99.4  & 76.3  & 99.5  & 75.9  & 99.4  & 75.7  & \textbf{99.1 } \\
          & DeepCNN & 74.2  & 86.8  & 73.7  & 95.0  & 73.4  & 95.5  & 72.7  & 96.9  & 73.4  & 97.5  & 74.4  & 96.3  & 73.9  & 94.9  & 73.6  & \textbf{96.0 } \\
          & ShallowCNN & 71.7  & 39.2  & 71.9  & 61.6  & 71.5  & 66.3  & 68.8  & 59.0  & 69.9  & 60.1  & 70.8  & 70.2  & 72.0  & 72.0  & 70.8  & \textbf{64.8 } \\
    \midrule
    \multirow{3}[2]{*}{Random pulse} & EEGNet & 76.1  & 98.3  & 77.1  & 99.4  & 76.8  & 99.2  & 76.5  & 99.6  & 76.9  & 99.6  & 77.2  & 99.7  & 77.0  & 99.6  & 76.9  & \textbf{99.5 } \\
          & DeepCNN & 73.0  & 95.1  & 74.9  & 97.4  & 74.5  & 97.3  & 73.7  & 98.8  & 73.3  & 99.1  & 74.2  & 98.1  & 74.4  & 97.4  & 74.2  & \textbf{98.0 } \\
          & ShallowCNN & 73.2  & 84.6  & 73.0  & 95.6  & 72.6  & 98.1  & 71.9  & 97.0  & 71.6  & 97.4  & 71.3  & 97.5  & 72.5  & 96.7  & 72.1  & \textbf{97.0 } \\
    \bottomrule
    \end{tabular} \normalsize
  \label{tab:trigger_MI1}%
\end{table}%

\begin{table}[htbp]
  \centering
  \caption{Classification and attack performances (\%) with poisoning rate 5\% and different backdoor triggers on MI2. `Average' was calculated excluding `Random'. Average ASRs higher than Random are marked in bold.} \setlength{\tabcolsep}{0.6mm} \fontsize{7pt}{\baselineskip}\selectfont
    \begin{tabular}{c|c|cc|cc|cc|cc|cc|cc|cc|cc}
    \toprule
    \multicolumn{1}{c|}{\multirow{2}[4]{*}{Trigger}} & \multirow{2}[4]{*}{Model} & \multicolumn{2}{c|}{Random} & \multicolumn{2}{c|}{MDS} & \multicolumn{2}{c|}{RDS} & \multicolumn{2}{c|}{MUS} & \multicolumn{2}{c|}{MMCS} & \multicolumn{2}{c|}{MUS+MDS} & \multicolumn{2}{c|}{MMCS+MDS} & \multicolumn{2}{c}{Average} \\
\cline{3-18}          &       & BCA   & ASR   & BCA   & ASR   & BCA   & ASR   & BCA   & ASR   & BCA   & ASR   & BCA   & ASR   & BCA   & ASR   & BCA   & ASR \\
    \midrule
    \multirow{3}[2]{*}{Sine wave} & EEGNet & 67.0  & 92.3  & 68.5  & 96.4  & 69.1  & 95.9  & 67.8  & 97.1  & 68.2  & 97.2  & 68.3  & 97.8  & 69.0  & 97.9  & 68.5  & \textbf{97.1 } \\
          & DeepCNN & 70.9  & 77.1  & 72.9  & 91.0  & 71.5  & 91.8  & 70.7  & 91.6  & 70.8  & 91.3  & 70.2  & 91.5  & 71.3  & 91.2  & 71.2  & \textbf{91.4 } \\
          & ShallowCNN & 72.6  & 79.5  & 72.4  & 96.8  & 71.7  & 98.6  & 71.7  & 96.9  & 71.6  & 98.0  & 71.5  & 96.9  & 72.3  & 96.5  & 71.9  & \textbf{97.3 } \\
    \midrule
    \multirow{3}[2]{*}{Sawtooth wave} & EEGNet & 67.0  & 87.5  & 67.5  & 94.4  & 67.6  & 94.5  & 66.8  & 95.0  & 66.8  & 95.9  & 67.4  & 96.8  & 67.7  & 96.3  & 67.3  & \textbf{95.5 } \\
          & DeepCNN & 70.4  & 74.5  & 72.6  & 88.2  & 72.5  & 90.1  & 70.9  & 89.3  & 71.3  & 90.9  & 71.9  & 88.6  & 73.6  & 79.3  & 72.1  & \textbf{87.7 } \\
          & ShallowCNN & 73.6  & 79.3  & 72.3  & 88.6  & 72.4  & 95.4  & 71.8  & 89.1  & 71.0  & 90.3  & 72.3  & 97.4  & 72.6  & 96.2  & 72.0  & \textbf{92.8 } \\
    \midrule
    \multirow{3}[2]{*}{Random pulse} & EEGNet & 67.1  & 94.7  & 68.0  & 97.8  & 67.6  & 97.7  & 66.9  & 98.0  & 67.4  & 97.7  & 68.1  & 98.1  & 71.0  & 84.9  & 68.2  & \textbf{95.7 } \\
          & DeepCNN & 71.0  & 84.9  & 73.1  & 92.2  & 72.4  & 93.5  & 71.1  & 95.2  & 72.0  & 95.1  & 71.9  & 93.7  & 72.2  & 93.1  & 72.1  & \textbf{93.8 } \\
          & ShallowCNN & 73.0  & 90.3  & 72.0  & 98.3  & 72.6  & 99.7  & 73.5  & 96.5  & 72.0  & 97.6  & 73.8  & 98.2  & 73.4  & 98.0  & 72.9  & \textbf{98.0 } \\
    \bottomrule
    \end{tabular}\normalsize
  \label{tab:trigger_MI2}%
\end{table}%

\subsection{More Challenging Scenarios}

\subsubsection{Fine-tuning}

Some labeled data from the target user may be used to fine-tune the trained (infected) model. It has been found that backdoor attacks can still be effective when only the last fully-connected layer is fine-tuned~\cite{gu2019badnets}. However, using the clean labeled data to retrain the entire model almost completely eliminated the backdoor in image classification~\cite{liu2018fine,liu2017neural}. We tested the attack performance of AP approaches under the challenging end-to-end fine-tuning scenario when 20\% samples from the target subject are labeled and 5\% source data are poisoned. The ASRs and BCAs of different AP approaches are shown in Table~\ref{tab:finetune}.

\begin{table}[htbp]   \centering
  \caption{Classification and attack performances (\%) with poisoning rate 5\% and labeling rate 20\% in fine-tuning. The best two ASRs for each model and each dataset are marked in bold.}   \setlength{\tabcolsep}{0.8mm} \fontsize{8pt}{\baselineskip}\selectfont
    \begin{tabular}{c|c|cc|cc|cc|cc|cc|cc|cc}
    \toprule
    \multirow{2}[4]{*}{Dataset} & \multirow{2}[4]{*}{Model} & \multicolumn{2}{c|}{Random} & \multicolumn{2}{c|}{MDS} & \multicolumn{2}{c|}{RDS} & \multicolumn{2}{c|}{MUS} & \multicolumn{2}{c|}{MMCS} & \multicolumn{2}{c|}{MUS+MDS} & \multicolumn{2}{c}{MMCS+MDS} \\
\cline{3-16}          &       & BCA   & ASR   & BCA   & ASR   & BCA   & ASR   & BCA   & ASR   & BCA   & ASR   & BCA   & ASR   & BCA   & ASR \\
    \midrule
    \multirow{3}[2]{*}{P300} & EEGNet & 62.6  & 11.0  & 64.8  & 13.3  & 65.1  & 16.1  & 63.7  & 19.6  & 63.7  & 21.2  & 65.0  & \textbf{22.0 } & 65.1  & \textbf{21.9 } \\
          & DeepCNN & 67.2  & 15.3  & 68.2  & 13.4  & 67.8  & 11.6  & 67.7  & \textbf{22.2 } & 67.6  & \textbf{19.0 } & 68.2  & 10.5  & 67.0  & 12.1  \\
          & ShallowCNN & 58.6  & 1.3   & 57.4  & 0.6   & 58.3  & 0.7   & 58.2  & 0.8   & 57.6  & 0.7   & 58.1  & 0.6   & 58.1  & 0.5  \\
    \midrule
    \multirow{3}[2]{*}{ERN} & EEGNet & 65.8  & 61.4  & 67.1  & 73.8  & 67.3  & 75.1  & 66.4  & 72.0  & 66.5  & 75.1  & 67.4  & \textbf{79.6 } & 67.0  & \textbf{80.1 } \\
          & DeepCNN & 66.1  & 36.5  & 66.0  & 58.0  & 65.8  & 63.9  & 65.4  & 57.7  & 65.6  & 58.5  & 65.6  & \textbf{67.7 } & 65.3  & \textbf{73.4 } \\
          & ShallowCNN & 62.9  & 15.0  & 66.0  & 46.4  & 66.3  & 52.2  & 66.1  & 38.5  & 65.1  & 46.2  & 65.7  & \textbf{61.4 } & \textbf{65.8 } & \textbf{59.6 } \\
    \midrule
    \multirow{3}[2]{*}{MI1} & EEGNet & 75.7  & 72.7  & 77.8  & 88.4  & 78.3  & 87.2  & 77.0  & 86.0  & 77.9  & \textbf{90.6 } & 78.1  & \textbf{90.6 } & 78.1  & 90.4  \\
          & DeepCNN & 71.1  & 12.8  & 75.4  & 39.4  & 73.3  & 34.3  & 72.7  & 29.3  & 72.7  & 34.0  & 74.9  & \textbf{47.0 } & 74.6  & \textbf{39.0 } \\
          & ShallowCNN & 73.4  & 0.9   & 75.7  & 7.1   & 75.4  & 9.4   & 75.0  & 4.2   & 75.2  & 5.0   & 75.5  & \textbf{15.7 } & 74.6  & \textbf{17.2 } \\
    \midrule
    \multirow{3}[2]{*}{MI2} & EEGNet & 77.3  & 45.6  & 78.6  & 61.3  & 79.0  & 65.8  & 78.7  & 75.0  & 79.1  & 80.1  & 78.6  & \textbf{80.7 } & 77.9  & \textbf{82.3 } \\
          & DeepCNN & 70.6  & 7.8   & 73.1  & 23.8  & 71.0  & 26.2  & 71.5  & 42.1  & 69.9  & 41.0  & 70.4  & \textbf{49.8 } & 70.6  & \textbf{50.2 } \\
          & ShallowCNN & 76.0  & 1.6   & 77.5  & 7.2   & 79.1  & 34.6  & 75.9  & 64.2  & 76.2  & 69.8  & 76.6  & \textbf{72.9 } & 77.6  & \textbf{71.8 } \\
    \midrule
    \multirow{3}[2]{*}{Average} & EEGNet & 70.3  & 47.7  & 72.1  & 59.2  & 72.4  & 61.1  & 71.5  & 63.1  & 71.8  & 66.7  & 72.2  & \textbf{68.2 } & 72.0  & \textbf{68.7 } \\
          & DeepCNN & 68.8  & 18.1  & 70.7  & 33.7  & 69.5  & 34.0  & 69.3  & 37.8  & 69.0  & 38.1  & 69.8  & \textbf{43.7 } & 69.4  & \textbf{43.7 } \\
          & ShallowCNN & 67.8  & 4.7   & 69.2  & 15.3  & 69.8  & 24.2  & 68.8  & 26.9  & 68.5  & 30.4  & 69.0  & \textbf{37.6 } & 69.0  & \textbf{37.2 } \\
    \bottomrule
    \end{tabular}  \normalsize
  \label{tab:finetune}%
\end{table}%

Table~\ref{tab:finetune} shows that:
\begin{enumerate}
\item The fine-tuning BCAs were generally higher than those in Table~\ref{tab:attack}, indicating backdoor attacks and AP approaches did not impact the classification performance. The ASRs were generally lower than those in Table~\ref{tab:attack}, consistent with the observations in image classification~\cite{liu2018fine,liu2017neural}, i.e., fine-tuning can defend against backdoor attacks to some extent.
\item Our proposed AP approaches still achieved higher ASRs than Random, and the combinational AP strategies generally had the best attack performance.
\item The ASRs dropped a lot, especially on P300 and MI1. The reason may be that the NPP amplitudes on these two datasets were small, and hence fine-tuning can more easily mask them.
\item Different models had different robustness to the same backdoor trigger on the same dataset, e.g., EEGNet had strong attack performance against fine-tuning.
\end{enumerate}

\begin{figure}[htbp]\centering
\subfloat[]{\includegraphics[width=\linewidth,clip]{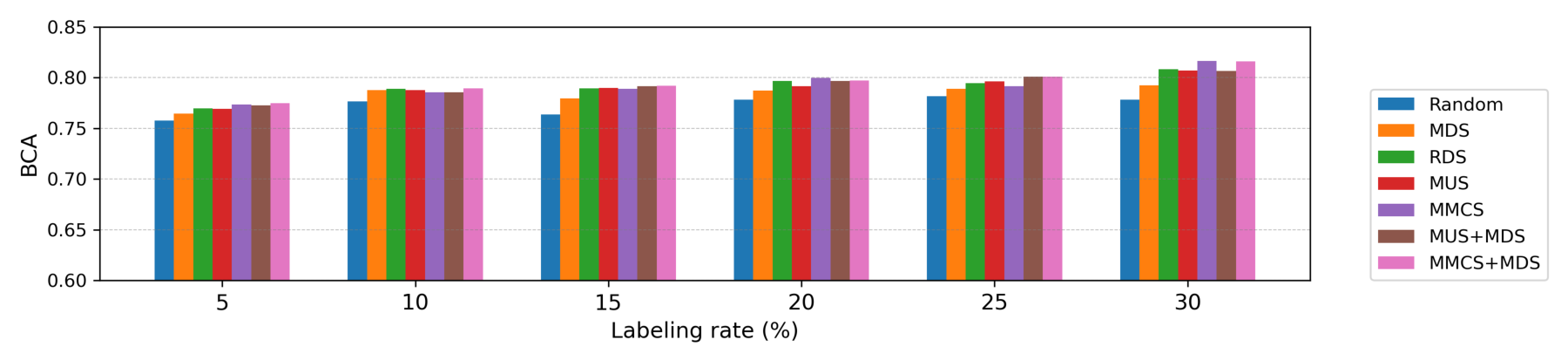}}\\
\subfloat[]{\includegraphics[width=\linewidth,clip]{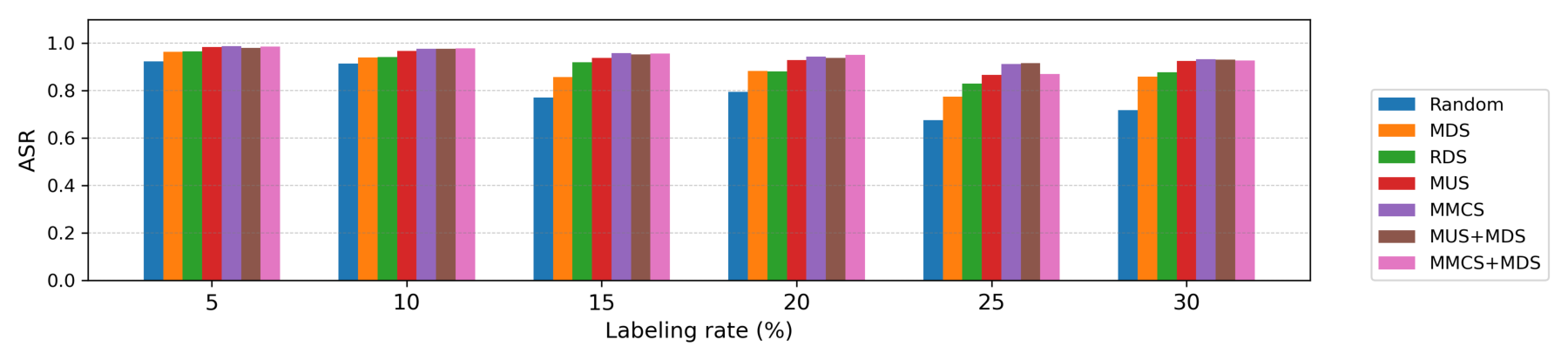}}
\caption{BCAs and ASRs when fine-tuned with different labeling rates on MI1 using EEGNet. a) BCAs; b) ASRs.} \label{fig:finetune}
\end{figure}

Figure~\ref{fig:finetune} shows the BCAs and ASRs of AP when fine-tuned with different labeling rates in the target domain on MI1 using EEGNet. Intuitively, the classification performance gradually improved with the increase of the number of clean labeled target-domain data. However, AP approaches still maintained good attack performance (high ASRs), outperforming Random.

\subsubsection{Data augmentation}

Fine-tuning on the clean data is a defense approach after training. Input preprocessing can be carried out during training to defend against backdoor attacks~\cite{borgnia2021strong,li2020rethinking}. Data augmentation on the raw EEG data is extensively used in model training to enhance the generalization ability of the model or to alleviate data insufficiency. Tables~\ref{tab:aug_P300}-\ref{tab:aug_MI2} show the results of using data augmentation strategies of \emph{noise, multiplication and frequency shift} in~\cite{freer2020data} and \emph{channel weakening} in~\cite{xia2022privacy} on the four datasets. The models were trained on the combination of the transformed data and the raw poisoned data, as in~\cite{freer2020data}.

Compared with the results in Table~\ref{tab:attack}, data augmentation strategies had no significant defense effect against backdoor attacks in general. The attack performances of the AP approaches were still better than Random, consistent with the above observations.

\begin{table}[htbp]
  \centering
  \caption{Classification and attack performances (\%) with poisoning rate 5\% and different data augmentation strategies on P300. `Average' was calculated excluding `Random'. Average ASRs higher than Random are marked in bold.}  \setlength{\tabcolsep}{0.6mm} \fontsize{7pt}{\baselineskip}\selectfont
    \begin{tabular}{c|c|cc|cc|cc|cc|cc|cc|cc|cc}
    \toprule
    \multicolumn{1}{c|}{\multirow{2}[4]{*}{Data augmentation}} & \multirow{2}[4]{*}{Model} & \multicolumn{2}{c|}{Random} & \multicolumn{2}{c|}{MDS} & \multicolumn{2}{c|}{RDS} & \multicolumn{2}{c|}{MUS} & \multicolumn{2}{c|}{MMCS} & \multicolumn{2}{c|}{MUS+MDS} & \multicolumn{2}{c|}{MMCS+MDS} & \multicolumn{2}{c}{Average} \\
\cline{3-18}          &       & BCA   & ASR   & BCA   & ASR   & BCA   & ASR   & BCA   & ASR   & BCA   & ASR   & BCA   & ASR   & BCA   & ASR   & BCA   & ASR \\
    \midrule
    \multirow{3}[2]{*}{Noise} & EEGNet & 62.0  & 81.6  & 62.3  & 70.6  & 62.1  & 80.2  & 61.6  & 92.4  & 61.4  & 91.4  & 61.9  & 87.9  & 61.8  & 89.8  & 61.8  & \textbf{85.4 } \\
          & DeepCNN & 62.9  & 68.7  & 63.0  & 68.9  & 63.0  & 74.9  & 62.7  & 88.0  & 62.8  & 86.5  & 63.0  & 86.4  & 63.4  & 83.7  & 63.0  & \textbf{81.4 } \\
          & ShallowCNN & 60.5  & 56.6  & 60.9  & 67.9  & 60.7  & 69.0  & 59.8  & 73.2  & 59.9  & 75.1  & 60.1  & 76.1  & 60.2  & 78.2  & 60.3  & \textbf{73.2 } \\
    \midrule
    \multirow{3}[2]{*}{Multiplication} & EEGNet & 62.8  & 77.2  & 62.4  & 65.9  & 62.5  & 77.2  & 62.6  & 88.1  & 62.3  & 89.7  & 62.2  & 83.9  & 62.0  & 85.9  & 62.3  & \textbf{81.8 } \\
          & DeepCNN & 62.7  & 69.7  & 63.5  & 73.4  & 63.6  & 77.1  & 63.2  & 87.5  & 63.2  & 86.7  & 63.4  & 88.7  & 63.2  & 85.7  & 63.3  & \textbf{83.2 } \\
          & ShallowCNN & 60.4  & 44.4  & 60.5  & 52.3  & 60.4  & 56.0  & 59.3  & 59.9  & 59.4  & 63.2  & 59.8  & 64.8  & 59.7  & 64.1  & 59.8  & \textbf{60.1 } \\
    \midrule
    \multirow{3}[2]{*}{Frequency shift} & EEGNet & 63.1  & 81.1  & 62.7  & 72.9  & 62.8  & 81.2  & 62.6  & 91.0  & 62.3  & 91.2  & 62.5  & 89.0  & 62.1  & 89.9  & 62.5  & \textbf{85.8 } \\
          & DeepCNN & 63.2  & 48.6  & 63.3  & 42.4  & 63.2  & 55.3  & 63.2  & 66.8  & 63.0  & 69.8  & 63.4  & 67.6  & 63.4  & 63.4  & 63.2  & \textbf{60.9 } \\
          & ShallowCNN & 60.4  & 54.9  & 60.3  & 63.5  & 60.5  & 65.9  & 59.4  & 69.6  & 59.4  & 72.0  & 59.8  & 73.8  & 59.7  & 73.4  & 59.9  & \textbf{69.7 } \\
    \midrule
    \multirow{3}[2]{*}{Channel weaken} & EEGNet & 62.4  & 80.3  & 62.5  & 71.6  & 62.6  & 78.5  & 62.1  & 91.4  & 62.3  & 91.3  & 62.1  & 90.0  & 62.0  & 92.7  & 62.3  & \textbf{85.9 } \\
          & DeepCNN & 62.8  & 81.4  & 63.3  & 81.8  & 62.9  & 85.9  & 62.7  & 95.3  & 62.6  & 95.0  & 63.0  & 95.6  & 62.7  & 95.1  & 62.9  & \textbf{91.4 } \\
          & ShallowCNN & 60.7  & 56.8  & 60.9  & 66.9  & 60.7  & 68.2  & 60.3  & 70.8  & 60.0  & 73.5  & 60.3  & 75.8  & 60.4  & 75.3  & 60.4  & \textbf{71.8 } \\
    \bottomrule
    \end{tabular}\normalsize
  \label{tab:aug_P300}%
\end{table}%

\begin{table}[htbp]
  \centering
  \caption{Classification and attack performances (\%) with poisoning rate 5\% and different data augmentation strategies on ERN. `Average' was calculated excluding `Random'. Average ASRs higher than Random are marked in bold.}   \setlength{\tabcolsep}{0.6mm} \fontsize{7pt}{\baselineskip}\selectfont
    \begin{tabular}{c|c|cc|cc|cc|cc|cc|cc|cc|cc}
    \toprule
    \multicolumn{1}{c|}{\multirow{2}[4]{*}{Data augmentation}} & \multirow{2}[4]{*}{Model} & \multicolumn{2}{c|}{Random} & \multicolumn{2}{c|}{MDS} & \multicolumn{2}{c|}{RDS} & \multicolumn{2}{c|}{MUS} & \multicolumn{2}{c|}{MMCS} & \multicolumn{2}{c|}{MUS+MDS} & \multicolumn{2}{c|}{MMCS+MDS} & \multicolumn{2}{c}{Average} \\
\cline{3-18}          &       & BCA   & ASR   & BCA   & ASR   & BCA   & ASR   & BCA   & ASR   & BCA   & ASR   & BCA   & ASR   & BCA   & ASR   & BCA   & ASR \\
    \midrule
    \multirow{3}[2]{*}{Noise} & EEGNet & 64.5  & 85.0  & 64.3  & 90.1  & 64.6  & 89.7  & 63.8  & 93.7  & 64.1  & 94.0  & 63.9  & 93.5  & 63.9  & 93.0  & 64.1  & \textbf{92.3 } \\
          & DeepCNN & 65.3  & 75.1  & 65.6  & 88.5  & 66.0  & 88.8  & 64.4  & 91.5  & 65.0  & 91.0  & 65.0  & 91.6  & 64.9  & 91.8  & 65.1  & \textbf{90.5 } \\
          & ShallowCNN & 64.2  & 46.4  & 63.6  & 73.5  & 63.7  & 76.4  & 63.6  & 72.7  & 63.3  & 73.5  & 63.4  & 80.9  & 63.1  & 82.8  & 63.5  & \textbf{76.6 } \\
    \midrule
    \multirow{3}[2]{*}{Multiplication} & EEGNet & 63.9  & 86.4  & 63.1  & 92.7  & 63.5  & 91.9  & 63.6  & 95.1  & 63.2  & 94.4  & 62.7  & 95.5  & 62.8  & 95.5  & 63.2  & \textbf{94.2 } \\
          & DeepCNN & 65.8  & 75.3  & 65.2  & 83.5  & 65.2  & 85.5  & 64.1  & 87.7  & 64.3  & 88.7  & 64.4  & 89.2  & 64.8  & 89.6  & 64.7  & \textbf{87.4 } \\
          & ShallowCNN & 64.5  & 46.8  & 63.1  & 69.4  & 63.5  & 71.1  & 62.7  & 69.3  & 62.4  & 69.2  & 62.2  & 75.8  & 62.1  & 76.3  & 62.7  & \textbf{71.9 } \\
    \midrule
    \multirow{3}[2]{*}{Frequency shift} & EEGNet & 59.7  & 73.4  & 62.5  & 87.2  & 63.0  & 85.1  & 62.7  & 88.3  & 62.6  & 88.4  & 62.3  & 89.8  & 62.1  & 89.6  & 62.5  & \textbf{88.1 } \\
          & DeepCNN & 65.5  & 74.1  & 65.3  & 85.7  & 65.6  & 86.0  & 64.4  & 86.3  & 64.4  & 85.8  & 64.8  & 89.3  & 63.8  & 86.9  & 64.7  & \textbf{86.7 } \\
          & ShallowCNN & 63.2  & 43.5  & 63.0  & 66.7  & 62.5  & 65.3  & 61.3  & 61.1  & 61.9  & 62.5  & 61.7  & 71.0  & 61.2  & 72.2  & 61.9  & \textbf{66.5 } \\
    \midrule
    \multirow{3}[2]{*}{Channel weaken} & EEGNet & 64.2  & 86.8  & 64.1  & 90.1  & 64.1  & 90.8  & 63.5  & 94.5  & 63.4  & 94.6  & 63.7  & 95.0  & 63.7  & 94.6  & 63.7  & \textbf{93.3 } \\
          & DeepCNN & 65.5  & 79.0  & 65.2  & 90.0  & 65.6  & 89.7  & 64.8  & 92.2  & 64.9  & 91.6  & 65.1  & 92.0  & 65.5  & 92.5  & 65.2  & \textbf{91.3 } \\
          & ShallowCNN & 63.7  & 44.5  & 63.3  & 72.4  & 63.6  & 72.9  & 62.2  & 68.5  & 62.4  & 69.4  & 62.3  & 78.9  & 62.7  & 78.3  & 62.7  & \textbf{73.4 } \\
    \bottomrule
    \end{tabular}\normalsize
  \label{tab:aug_ERN}%
\end{table}%

\begin{table}[htbp]
  \centering
  \caption{classification and attack performances (\%) with poisoning rate 5\% and different data augmentation strategies on MI1. `Average' was calculated excluding `Random'. Average ASRs higher than Random are marked in bold.}  \setlength{\tabcolsep}{0.6mm} \fontsize{7pt}{\baselineskip}\selectfont
    \begin{tabular}{c|c|cc|cc|cc|cc|cc|cc|cc|cc}
    \toprule
    \multicolumn{1}{c|}{\multirow{2}[4]{*}{Data augmentation}} & \multirow{2}[4]{*}{Model} & \multicolumn{2}{c|}{Random} & \multicolumn{2}{c|}{MDS} & \multicolumn{2}{c|}{RDS} & \multicolumn{2}{c|}{MUS} & \multicolumn{2}{c|}{MMCS} & \multicolumn{2}{c|}{MUS+MDS} & \multicolumn{2}{c|}{MMCS+MDS} & \multicolumn{2}{c}{Average} \\
\cline{3-18}          &       & BCA   & ASR   & BCA   & ASR   & BCA   & ASR   & BCA   & ASR   & BCA   & ASR   & BCA   & ASR   & BCA   & ASR   & BCA   & ASR \\
    \midrule
    \multirow{3}[2]{*}{Noise} & EEGNet & 75.6  & 91.0  & 75.7  & 96.2  & 75.0  & 95.8  & 74.3  & 98.2  & 74.4  & 98.2  & 74.9  & 98.1  & 74.6  & 97.9  & 74.8  & \textbf{97.4 } \\
          & DeepCNN & 73.1  & 69.5  & 74.1  & 84.1  & 74.5  & 87.4  & 72.7  & 89.2  & 72.6  & 88.2  & 74.5  & 87.5  & 74.2  & 86.3  & 73.7  & \textbf{87.1 } \\
          & ShallowCNN & 71.7  & 0.6   & 72.5  & 4.1   & 71.5  & 12.7  & 67.9  & 1.2   & 68.3  & 1.7   & 71.6  & 43.0  & 72.1  & 44.8  & 70.7  & \textbf{17.9 } \\
    \midrule
    \multirow{3}[2]{*}{Multiplication} & EEGNet & 76.0  & 95.7  & 75.8  & 97.8  & 75.2  & 97.4  & 75.1  & 99.3  & 75.2  & 99.1  & 75.4  & 98.7  & 72.9  & 74.5  & 74.9  & 94.5  \\
          & DeepCNN & 72.9  & 74.5  & 72.7  & 86.8  & 72.2  & 90.7  & 72.0  & 90.6  & 72.5  & 91.5  & 72.1  & 90.7  & 71.7  & 87.7  & 72.2  & \textbf{89.7 } \\
          & ShallowCNN & 70.2  & 0.7   & 70.8  & 3.3   & 70.2  & 12.5  & 67.8  & 1.7   & 67.3  & 1.7   & 69.6  & 39.0  & 69.6  & 41.3  & 69.2  & \textbf{16.6 } \\
    \midrule
    \multirow{3}[2]{*}{Frequency shift} & EEGNet & 75.5  & 95.0  & 76.2  & 99.0  & 76.1  & 98.3  & 75.4  & 99.3  & 75.3  & 99.6  & 75.2  & 99.5  & 75.4  & 99.2  & 75.6  & \textbf{99.1 } \\
          & DeepCNN & 73.1  & 34.4  & 73.5  & 49.4  & 73.2  & 58.2  & 70.2  & 55.0  & 70.2  & 51.4  & 72.1  & 69.9  & 73.1  & 64.6  & 72.0  & \textbf{58.1 } \\
          & ShallowCNN & 70.7  & 0.4   & 70.9  & 1.2   & 70.1  & 5.6   & 68.4  & 0.6   & 68.1  & 0.5   & 69.4  & 18.8  & 70.4  & 22.3  & 69.5  & \textbf{8.2 } \\
    \midrule
    \multirow{3}[2]{*}{Channel weaken} & EEGNet & 76.5  & 97.0  & 76.8  & 98.4  & 76.8  & 98.2  & 75.9  & 99.3  & 76.5  & 99.3  & 76.7  & 99.1  & 76.4  & 98.9  & 76.5  & \textbf{98.9 } \\
          & DeepCNN & 74.4  & 76.7  & 74.2  & 89.9  & 74.6  & 94.0  & 73.6  & 93.9  & 73.7  & 93.8  & 74.6  & 94.6  & 74.5  & 92.8  & 74.2  & \textbf{93.2 } \\
          & ShallowCNN & 71.1  & 1.3   & 70.7  & 9.6   & 70.1  & 27.7  & 68.4  & 3.2   & 68.4  & 3.8   & 68.4  & 63.9  & 68.4  & 64.9  & 69.1  & \textbf{28.9 } \\
    \bottomrule
    \end{tabular}\normalsize
  \label{tab:aug_MI1}%
\end{table}%

\begin{table}[htbp]
  \centering
  \caption{classification and attack performances (\%) with poisoning rate 5\% and different data augmentation strategies on MI2. `Average' was calculated excluding `Random'. Average ASRs higher than Random are marked in bold.}   \setlength{\tabcolsep}{0.6mm} \fontsize{7pt}{\baselineskip}\selectfont
    \begin{tabular}{c|c|cc|cc|cc|cc|cc|cc|cc|cc}
    \toprule
    \multicolumn{1}{c|}{\multirow{2}[4]{*}{Data augmentation}} & \multirow{2}[4]{*}{Model} & \multicolumn{2}{c|}{Random} & \multicolumn{2}{c|}{MDS} & \multicolumn{2}{c|}{RDS} & \multicolumn{2}{c|}{MUS} & \multicolumn{2}{c|}{MMCS} & \multicolumn{2}{c|}{MUS+MDS} & \multicolumn{2}{c|}{MMCS+MDS} & \multicolumn{2}{c}{Average} \\
\cline{3-18}          &       & BCA   & ASR   & BCA   & ASR   & BCA   & ASR   & BCA   & ASR   & BCA   & ASR   & BCA   & ASR   & BCA   & ASR   & BCA   & ASR \\
    \midrule
    \multirow{3}[2]{*}{Noise} & EEGNet & 67.7  & 68.6  & 69.6  & 85.1  & 69.9  & 85.9  & 68.8  & 84.8  & 68.8  & 86.0  & 69.2  & 88.5  & 69.1  & 88.0  & 69.2  & \textbf{86.4 } \\
          & DeepCNN & 70.2  & 54.8  & 71.0  & 81.6  & 71.0  & 85.7  & 70.7  & 84.0  & 70.3  & 83.0  & 71.8  & 85.3  & 71.9  & 61.8  & 71.1  & \textbf{80.2 } \\
          & ShallowCNN & 71.9  & 61.8  & 72.0  & 86.9  & 71.5  & 91.7  & 71.3  & 84.6  & 71.0  & 88.2  & 71.6  & 93.3  & 71.3  & 93.3  & 71.4  & \textbf{89.7 } \\
    \midrule
    \multirow{3}[2]{*}{Multiplication} & EEGNet & 67.8  & 86.4  & 69.4  & 94.0  & 69.3  & 94.5  & 68.6  & 93.8  & 69.4  & 94.3  & 69.1  & 96.0  & 68.5  & 95.7  & 69.0  & \textbf{94.7 } \\
          & DeepCNN & 71.4  & 66.1  & 72.3  & 89.5  & 72.8  & 91.9  & 71.0  & 91.9  & 70.4  & 90.6  & 72.6  & 91.2  & 71.2  & 70.3  & 71.7  & \textbf{87.6 } \\
          & ShallowCNN & 71.2  & 70.3  & 72.0  & 86.3  & 71.0  & 96.3  & 70.3  & 87.7  & 70.1  & 88.8  & 71.6  & 94.4  & 71.2  & 94.2  & 71.0  & \textbf{91.3 } \\
    \midrule
    \multirow{3}[2]{*}{Frequency shift} & EEGNet & 68.8  & 74.6  & 69.5  & 91.4  & 69.5  & 92.3  & 68.9  & 87.9  & 69.2  & 91.7  & 69.3  & 93.7  & 69.0  & 93.5  & 69.2  & \textbf{91.7 } \\
          & DeepCNN & 71.9  & 59.6  & 71.8  & 89.9  & 71.9  & 91.9  & 71.5  & 88.5  & 71.5  & 89.6  & 70.9  & 92.1  & 71.7  & 90.0  & 71.6  & \textbf{90.3 } \\
          & ShallowCNN & 70.2  & 68.1  & 70.7  & 77.8  & 71.1  & 85.2  & 68.5  & 76.8  & 69.5  & 77.3  & 70.3  & 95.2  & 69.9  & 92.9  & 70.0  & \textbf{84.2 } \\
    \midrule
    \multirow{3}[2]{*}{Channel weaken} & EEGNet & 68.6  & 83.4  & 68.6  & 83.4  & 69.1  & 84.0  & 67.8  & 81.8  & 68.3  & 84.1  & 68.1  & 86.2  & 68.3  & 85.8  & 68.4  & \textbf{84.2 } \\
          & DeepCNN & 73.0  & 57.5  & 72.5  & 84.5  & 71.9  & 90.1  & 71.4  & 86.6  & 71.0  & 88.5  & 71.6  & 89.3  & 72.1  & 89.8  & 71.7  & \textbf{88.1 } \\
          & ShallowCNN & 71.6  & 65.5  & 71.0  & 85.6  & 71.1  & 96.5  & 69.7  & 81.4  & 69.6  & 83.9  & 70.6  & 96.1  & 71.0  & 96.3  & 70.5  & \textbf{90.0 } \\
    \bottomrule
    \end{tabular} \normalsize
  \label{tab:aug_MI2}%
\end{table}%

\subsubsection{Simultaneous cross-subject and cross-task TL}

Cross-task TL is a challenging scenario in EEG-based BCIs, where the labeled data from other similar tasks (source domains) are used to improve the calibration for a new task (target domain)~\cite{wu2020transfer}. We consider the more challenging simultaneous cross-subject and cross-task TL scenario, where cross-task means transferring between different label spaces. Specifically, the label space of the target subject is different from that of the source subjects, e.g., the source data of `left-hand' and `right-hand' MIs may be used to calibrate `feet' and `tongue' MIs of the target subject.

In AP attacks, the target label specified by the attacker must be from the target label space. Therefore, we considered the scenario that the label spaces of the source and target subjects are partially different.

\paragraph{Cross-task TL}
We used label alignment (LA)~\cite{he2020different} to align the source data to the target data. Assume the target subject and the source subjects have the same number of classes $M$, but their class labels are partially different. The goal of LA is to transform the trials of the $m$-th ($m=1,2,\cdots,M$) class from the $s$-th ($s=1,2,\cdots,S$) source subject with a matrix $A_{s,m}$, so as to minimize the distance of the mean covariance matrix from that in the $m$-th class of the target subject, i.e.,
\begin{eqnarray}
A_{s,m} = \mathop{\arg\min}\limits_{A}\|A\bar{R}_{s,m}A^T-\bar{R}_{t,m}\|_F^2,\quad m=1,2,\cdots,M;\ s=1,2,\cdots,S,
\end{eqnarray}
where $\bar{R}_{s,m}$ is the mean covariance matrix of the $m$-th class of the $s$-th source subject, and $\bar{R}_{t,m}$ the mean covariance matrix of the corresponding target class.

The transformation matrix for the $m$-th class of the $s$-th source subject is then
\begin{eqnarray}
A_{s,m} = \bar{R}_{t,m}^\frac{1}{2} \bar{R}_{s,m}^{-\frac{1}{2}}.
\end{eqnarray}

$\bar{R}_{t,m}$ requires some label information in target subject, which can be solved by selecting a small number of target samples by $k$-means clustering based on Riemannian distance for labeling.

Finally, the $n$-th trial in the $m$-th class of the $s$-th source subject is transformed to
\begin{eqnarray}
\hat{X}_{s,m}^n=A_{s,m}X_{s,m}^n,\quad n=1,...,N_s.
\end{eqnarray}

As we consider partially different label spaces between the source and target subjects, we match the label of each source subject with the same label of the target subject, and then randomly match each remaining source label with a remaining target label, as in~\cite{he2020different}.

\paragraph{Results}
We performed leave-one-subject-out cross-validation on the three classes (`left-hand', `right-hand' and `tongue') of MI1 for cross-subject evaluation. The dataset was further divided into a source sub-dataset that had the two classes of `left-hand' and `right-hand' and a target sub-dataset that had the two classes of `tongue' and `right-hand' for cross-task evaluation using LA. We set $k=10$ in $k$-means clustering of LA, and the target label for AP attacks as `right-hand'. All other parameters were the same as those in Section~\ref{sec:ap_attack}. The average results of five repetitions are shown in Table~\ref{tab:LA}.

\begin{table}[htbp]
  \centering
  \caption{Classification and attack performances (\%) with poisoning rate 20\% in simultaneous cross-subject and cross-task TL scenario on the three-class MI1 dataset. `Average' was calculated excluding `Random'. Average ASRs higher than Random are marked in bold.}  \setlength{\tabcolsep}{1mm} \fontsize{8pt}{\baselineskip}\selectfont
    \begin{tabular}{c|cc|cc|cc|cc|cc|cc|cc|cc}
    \toprule
    \multirow{2}[4]{*}{Model} & \multicolumn{2}{c|}{Random} & \multicolumn{2}{c|}{MDS} & \multicolumn{2}{c|}{RDS} & \multicolumn{2}{c|}{MUS} & \multicolumn{2}{c|}{MMCS} & \multicolumn{2}{c|}{MUS+MDS} & \multicolumn{2}{c|}{MMCS+MDS} & \multicolumn{2}{c}{Average} \\
\cmidrule{2-17}          & BCA   & ASR   & BCA   & ASR   & BCA   & ASR   & BCA   & ASR   & BCA   & ASR   & BCA   & ASR   & BCA   & ASR   & BCA   & ASR \\
    \midrule
    EEGNet & 70.9  & 24.7  & 69.7  & 28.3  & 70.6  & 29.1  & 70.4  & 25.9  & 70.8  & 27.6  & 71.5  & 24.1  & 70.3  & 20.3  & 70.5  & \textbf{25.9 } \\
    DeepCNN & 70.3  & 20.3  & 69.7  & 21.2  & 70.6  & 20.7  & 69.6  & 25.7  & 69.8  & 21.7  & 69.4  & 22.0  & 71.3  & 19.1  & 70.0  & \textbf{21.8 } \\
    ShallowCNN & 74.2  & 6.5   & 70.8  & 11.3  & 73.6  & 6.6   & 72.0  & 5.2   & 72.4  & 5.8   & 73.8  & 7.7   & 74.1  & 7.7   & 72.8  & \textbf{7.4 } \\
    \bottomrule
    \end{tabular}\normalsize
  \label{tab:LA}%
\end{table}%

The classification tasks were still successful (the BCAs were well above the 50\% chance level for binary classification), when the source subjects and the target subject had different label spaces, indicating the effectiveness of LA. AP approaches slightly outperformed Random in most cases, but the ASRs were much lower than those in previous experiments. This may be because LA transforms each class by a different matrix, and hence the actual backdoor is distorted (much different from the original backdoor).
%

\subsection{Computational cost}

Table~\ref{tab:cost} shows the computational cost (seconds) of different AP approaches on the four datasets using EEGNet as the target model, averaged over different subjects. It includes the time of two stages: the attacker generates the backdoor trigger on the source data, and tests the infected model on the target subject. Generally, the computational costs of our proposed AP approaches and Random selection are comparable. Due to the use of early stopping, the results on different datasets were sometimes inconsistent.

\begin{table}[htbp]
  \centering
  \caption{Computational cost (seconds) of different AP approaches on the four datasets using EEGNet as the target model, running on a single GeForce GTX 1080 GPU.}\setlength{\tabcolsep}{4mm} \fontsize{8pt}{\baselineskip}\selectfont
    \begin{tabular}{cccccccc}
    \toprule
          & Random & MDS   & RDS   & MUS   & MMCS  & MUS+MDS & MMCS+MDS \\
    \midrule
    P300  & 111.79  & 129.95  & 142.77  & 181.53  & 120.65  & 134.93  & 128.55  \\
    ERN   & 49.66  & 103.98  & 95.07  & 102.37  & 76.97  & 70.68  & 69.03  \\
    MI1   & 63.57  & 107.56  & 100.14  & 73.60  & 105.96  & 80.37  & 78.55  \\
    MI2   & 33.54  & 66.03  & 73.97  & 64.58  & 66.12  & 62.10  & 60.77 \\
    \bottomrule
    \end{tabular}\normalsize
  \label{tab:cost}%
\end{table}%

\subsection{AP attacks on the SVM classifier}

To our knowledge, no backdoor attack approach has been proposed for traditional classifiers. We tested our AP approaches on the SVM classifier. Specifically, the same trigger settings as in Section~\ref{sec:ap_attack} on the CNN models were applied. xDAWN spatial filtering~\cite{xDAWN} and SVM classifier were used on P300 and ERN, and CSP filtering~\cite{CSP} and SVM on MI1 and MI2. The results are shown in Table~\ref{tab:SVM}.

\begin{table}[htbp]
  \centering
  \caption{Classification and attack performances (\%) with poisoning rate 5\% on traditional models. The best two ASRs are marked in bold.}  \setlength{\tabcolsep}{0.8mm} \fontsize{7pt}{\baselineskip}\selectfont
    \begin{tabular}{c|c|cc|cc|cc|cc|cc|cc|cc|cc}
    \toprule
    \multirow{3}[6]{*}{Dataset} & \multirow{3}[6]{*}{Model} & \multicolumn{2}{c|}{Baseline} & \multicolumn{14}{c}{Active Poisoning} \\
\cline{3-18}          &       & \multirow{2}[4]{*}{BCA} & \multirow{2}[4]{*}{ASR} & \multicolumn{2}{c|}{Random} & \multicolumn{2}{c|}{MDS} & \multicolumn{2}{c|}{RDS} & \multicolumn{2}{c|}{MUS} & \multicolumn{2}{c|}{MMCS} & \multicolumn{2}{c|}{MUS+MDS} & \multicolumn{2}{c}{MMCS+MDS} \\
\cline{5-18}          &       &       &       & BCA   & ASR   & BCA   & ASR   & BCA   & ASR   & BCA   & ASR   & BCA   & ASR   & BCA   & ASR   & BCA   & ASR \\
    \midrule
    P300  & xDAWN+SVM & 58.3  & 13.5  & 57.1  & 91.8  & 57.9  & 73.1  & 57.1  & 90.4  & 56.3  & \textbf{95.0 } & 56.7  & \textbf{95.1 } & 56.4  & 94.1  & 56.6  & 91.3  \\
    ERN   & xDAWN+SVM & 65.1  & 9.1   & 64.7  & 32.4  & 64.5  & 29.0  & 63.9  & 36.4  & 62.0  & \textbf{41.4 } & 61.9  & 40.6  & 62.5  & \textbf{41.2 } & 63.4  & 30.7  \\
    MI1   & CSP+SVM & 71.6  & 0.0   & 71.5  & 0.0   & 67.8  & 0.3   & 71.9  & 0.0   & 68.5  & 0.0   & 65.7  & 0.2   & 67.8  & \textbf{0.4 } & 66.6  & 0.0  \\
    MI2   & CSP+SVM & 79.5  & 0.0   & 79.0  & 0.2   & 65.6  & 0.0   & 66.2  & 0.0   & 71.5  & 0.0   & 72.6  & 0.0   & 69.1  & 0.0   & 68.0  & 0.0  \\
    Average & -     & 68.6  & 5.7   & 68.1  & 31.1  & 64.0  & 25.6  & 64.8  & 31.7  & 64.6  & \textbf{34.1 } & 64.2  & \textbf{34.0 } & 64.0  & 33.9  & 63.6  & 30.5  \\
    \bottomrule
    \end{tabular}\normalsize
  \label{tab:SVM}%
\end{table}%

`CSP+SVM' model for MI had strong resistance to backdoor attacks, resulted in nearly zero ASRs. Attacks on `xDAWN+SVM' model for P300 and ERN were still effective. The model-based AP approaches (MUS and MMCS) achieved better attack performance (higher ASRs) than Random, as for CNN models. However, the diversity-based AP approach, MDS, was ineffective in attacking the traditional models. Our future research will try to improve it.

\section{Conclusions} \label{sect:Conclusion}

TL has been widely used in EEG-based BCIs for reducing calibration efforts. However, backdoor attacks could be introduced through TL. Accordingly, this study explored backdoor attacks in TL of EEG-based BCIs, where source-domain data are poisoned by an NPP trigger and then used in TL. We verified that the classification performance remains good on benign target-domain samples, but once the trigger is injected, the attacked samples would be misclassified into an attacker-specified target class with a very high probability. We have proposed several AP approaches to select source-domain samples that are most effective in embedding the backdoor pattern to improve the attack success rate and efficiency. Experiments on four EEG datasets and three CNN models demonstrated the success of backdoor attacks in TL scenarios and the effectiveness of our proposed AP approaches.

To our knowledge, this is the first study on backdoor attacks on TL models in EEG-based BCIs. It exposes a serious security risk in BCIs, which will be addressed in our future research.

\Acknowledgements{This work was supported by the Open Research Projects of Zhejiang Lab (2021KE0AB04), Technology Innovation Project of Hubei Province of China (2019AEA171), and Hubei Province Funds for Distinguished Young Scholars (2020CFA050).}

\end{document}